\documentclass[superscriptaddress,nobibnotes,amsmath,amssymb,notitlepage,twocolumn,pra,longbibliography]{revtex4-2}

\usepackage{bm,braket}
\usepackage[toc,page]{appendix}
\usepackage{comment}
\usepackage{dcolumn}

\usepackage{amsfonts}

\usepackage{graphicx,color,hyperref}
\usepackage[caption=false]{subfig}
\hypersetup{colorlinks=true, linkcolor=blue, citecolor=blue, urlcolor=blue} 

\graphicspath{{./img/}}

\newcommand{\beq}[1]{\begin{equation}\label{#1}}
\newcommand{\eep}{\;.\end{equation}}
\newcommand{\eec}{\;,\end{equation}}
\newcommand{\eeq}{\end{equation}}

\newcommand*\dd{\mathop{}\!\mathrm{d}} 

\newcommand{\om}{\omega}

\DeclareMathAlphabet{\mathcal}{OMS}{cmsy}{m}{n} 

\renewcommand{\vec}[1]{{\bf #1}}

\newcommand{\kv}{\vec{k}}

\newcommand{\rv}{\vec{r}}

\usepackage{amsmath}
\usepackage{amssymb}
\usepackage{xcolor}
\usepackage{bbm}
\usepackage{physics}
\usepackage{float}
\usepackage{graphicx}
\usepackage{dcolumn} 
\usepackage{bm} 
\usepackage{siunitx}
\usepackage{enumitem}  

\makeatletter
\renewcommand*{\fnum@figure}{{\normalfont\bfseries \figurename~\thefigure}}
\makeatother

\allowdisplaybreaks

\definecolor{orange}{rgb}{1,0.5,0}

\newcommand{\sect}[1]{\vspace{0.3em}{\it #1.}---}

\graphicspath{{./img/}}

\DeclareMathAlphabet{\mathcal}{OMS}{cmsy}{m}{n} 

\newcommand{\ii}{\mathrm{i}}

\makeatletter
\newcommand{\specificthanks}[1]{\@fnsymbol{#1}}
\makeatother

\begin{document}

\preprint{APS/123-QED}

\title{Quantized Integrated Shift Effect in Multigap Topological Phases}

\author{Wojciech J. Jankowski}
\email{wjj25@cam.ac.uk}
\affiliation{TCM Group, Cavendish Laboratory, Department of Physics, J J Thomson Avenue, Cambridge CB3 0HE, United Kingdom}

\author{Robert-Jan Slager}
\email{rjs269@cam.ac.uk}
\affiliation{TCM Group, Cavendish Laboratory, Department of Physics, J J Thomson Avenue, Cambridge CB3 0HE, United Kingdom}

\date{\today}

\begin{abstract}
    We show that certain three-dimensional multigap topological insulators can host quantized integrated shift photoconductivities due to bulk invariants that are defined under reality conditions imposed by additional symmetries. We recast the quantization in terms of the integrated torsion tensor and the non-Abelian Berry connection constituting Chern-Simons forms. Physically, we recognize that the topological quantization emerges purely from virtual transitions contributing to the optical response. Our findings provide another quantized electromagnetic dc response due to the nontrivial band topology, beyond the quantum anomalous Hall effect of Chern insulators and quantized circular photogalvanic effect found in Weyl semimetals.   
\end{abstract} 

\maketitle

\sect{Introduction}
Topological insulators and semimetals constitute an active research area in condensed matter physics~\cite{Rmp1,Rmp2, Weylrmp, Kitaev}. Amid a fast track of progress, the past years have seen the emergence of a rather uniform view on a large fraction of topological materials upon including crystalline symmetries~\cite{Clas1,clas2,clas3,Clas4,Codefects2,Clas5}. In particular, considering how irreducible representations at high symmetry points need to match over the whole Brillouin zone, one may obtain a set of consistency equations~\cite{clas3} that upon comparing to real space Wannierization conditions leverages universal schemes to discern topological materials~\cite{Clas4,Clas5}.

More recently, however, further advances have revealed a new class of topological phases that depend on multigap conditions~\cite{bouhon2020geometric,  bouhon2019nonabelian, BJY_nielsen, bouhon2022multigap, bouhon2018wilson}, i.e.,~their band topology is protected by the presence of multiple band gaps, and generally cannot be  captured by symmetry-indicated methods~\cite{clas3, Clas4,Clas5}. In such phases, band spaces can acquire novel multigap invariants. In particular, in the presence of a reality condition, i.e., Hamiltonian being representable as a real symmetric matrix, ensured by the presence of $\mathcal{C}_2\mathcal{T}$ (twofold rotations and time reversal) or $\mathcal{P}\mathcal{T}$ (parity and time reversal) symmetries, band degeneracies can carry non-Abelian charges~\cite{Wu1273}, akin to disclination charges in a biaxial nematic~\cite{volovik2018investigation, Beekman20171, PRX2016, Kamienrmp}. As a result of braiding processes, band subspaces can subsequently come to host a set of similarly valued charges that cannot be removed, posing an obstruction to annihilate as long as the gaps with the other parts of the spectrum are maintained~\cite{BJY_nielsen, bouhon2019nonabelian,Jiang2021,jankowski2023disorderinduced}. This obstruction to annihilate is quantified by a multigap invariant, the Euler class~\cite{bouhon2018wilson, BJY_nielsen, Ahn2019, bouhon2019nonabelian, bouhon2020geometric,Jiang2021} evaluated over Brillouin zone patches, to discern whether the band nodes within the two-band subspace (i.e.,~that pair of bands) are stable to being annihilated~\cite{BJY_nielsen,Jiang2021}. 

While the multigap point of view opens up novel theoretical pursuits~\cite{wahl2024exactprojectedentangledpair}, utilizing relations with \mbox{homotopy} theory~\cite{bouhon2020geometric}, recent developments have culminated \mbox{physical signatures~\cite{jankowski2023optical}. In the out-of-equilibrium} context, novel Floquet phases were postulated~\cite{slager2022floquet}. In~addition,  monopole--antimonopole creation in the quench dynamics of Euler phases was predicted~\cite{Unal_quenched_Euler} and recently observed~\cite{zhao2022observation}. Similarly, stress or strain has been predicted to lead to braiding and Euler invariants in the context of phononic~\cite{Peng2021, Peng2022Multi} and electronic spectra~\cite{bouhon2019nonabelian, magnetic} of real materials, where in the electronic case it was also shown that  multigap physics can arise by virtue of \mbox{structural} phase transitions~\cite{chen2021manipulation}. Finally, multigap topologies are increasingly implemented in metamaterials~\cite{Guo1Dexp, jiang_meron, Jiang2021, qiu2022minimal, yang2023nonabelian}.

\begin{figure}
\centering
  \includegraphics[width=1.0\columnwidth]{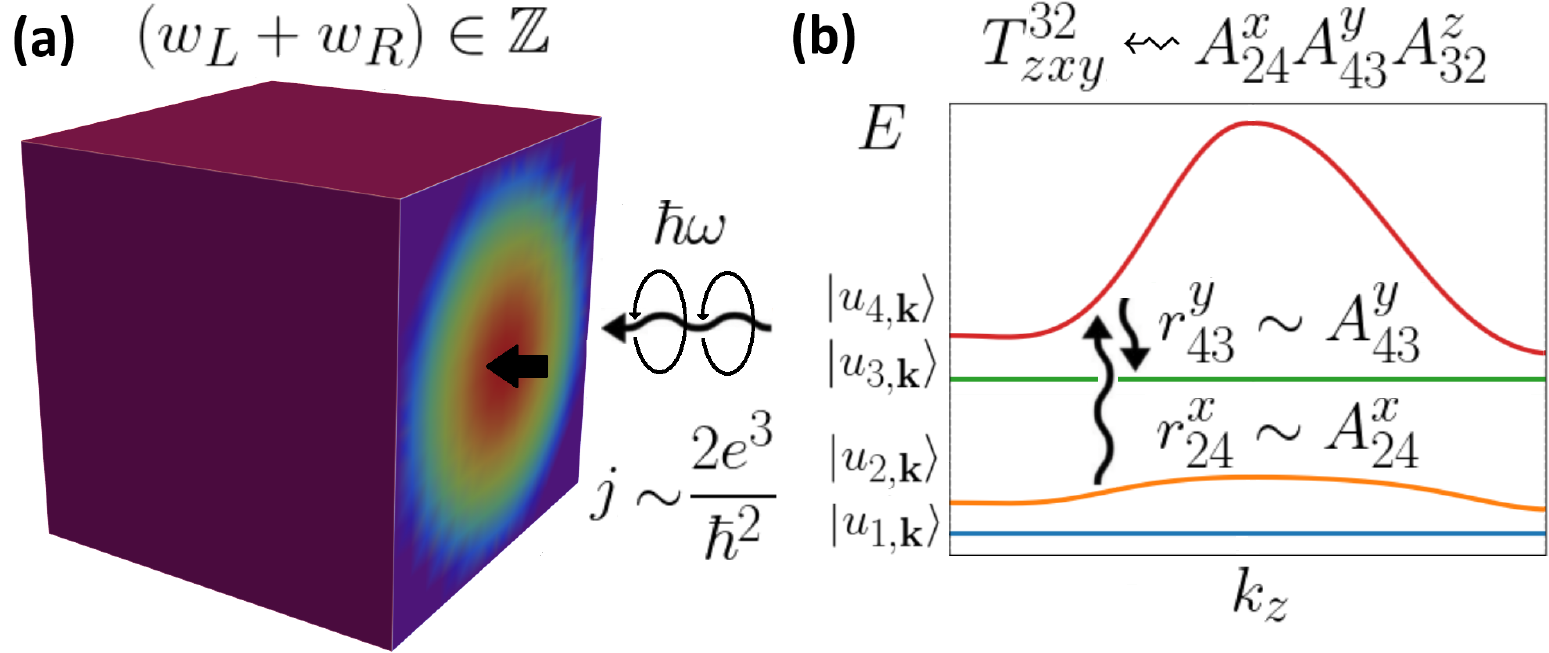}
  \caption{ \textbf{(a)} Quantized bulk photovoltaic shift currents induced by incident light. A second-order dc current density response $j_z(0)$ of a multigap topological insulator with summed photoconductivities quantized in terms of $2\frac{e^3}{\hbar^2}$, on coupling circularly polarized light with photon energies $\hbar\om$ and electric field $\mathcal{E}(\frac{\hat{x} + \text{i}\hat{y}}{\sqrt{2}})$. The shift response is due to virtual transitions captured by the torsion tensor $T^{mn}_{abc}$ between the occupied and unoccupied bands. \textbf{(b)} Diagrammatic representation of the corresponding virtual transitions. Transition dipole amplitudes $r^x_{24}, r^y_{43}$ from band $\ket{u_{2,\kv}}$ to $\ket{u_{3,\kv}}$, virtually through $\ket{u_{4,\kv}}$, are indicated. These are proportional to the elements of non-Abelian Berry connection $A^x_{24}$, $A^y_{43}$, contributing to the torsion tensor element $T^{32}_{zxy}$.}
\label{fig:schematics}
\end{figure}

A hallmark of topological materials is that they can render quantized responses. Chern bands, for example, were shown to exhibit quantized anomalous Hall (QAH) and quantized dichroism responses~\cite{PhysRevLett.61.2015, PhysRevLett.103.116803, doi:10.1126/sciadv.1701207, Asteria_2019}. Similarly, Weyl semimetals can be probed by second-order quantized circular photogalvanic effects (QCPGE)~\cite{de_Juan_2017, Parker_2019, Avdoshkin_2020, OrensteinRev}. We note that these phenomena are presently more and more understood from the upcoming perspective of quantum geometry. Following formulations of quantum geometric tensors (QGTs) quite some time ago~\cite{provost1980riemannian, resta_2011_metric}, these concepts have recently been reinvigorated in several contexts that range from superconductivity and superfluidity to transport properties~\cite{peotta2015superfluidity,PhysRevLett.124.167002,doi:10.1073/pnas.2106744118,PhysRevB.105.085154}. Response theories fit particularly well in this context~\cite{tormaessay} as they relate directly to elements of QGTs (see also SM~\cite{SI}). Recent progress includes rephrasing matrix elements of optical processes in terms of Riemannian geometry~\cite{Ahn_2021rio}, and, moreover, a universal strategy in terms of Pl\"ucker embeddings~\cite{bouhon2023quantum}, to calculate geometric entities and QGTs, and thus accordingly optical responses, in general multiband systems. 

The mentioned mathematical Pl\"ucker framework works in two directions. Apart from quantifying geometry, it allows for an efficient method to model topological phases for arbitrarily partitioned band spaces. As such, this approach has stood at the basis of generalizing multigap topologies to fully generalized settings, as it allows for a direct evaluation of the classifying space set by the Hamiltonian~\cite{bouhon2020geometric, bouhon2022multigap}. Particular examples include three dimensional (3D) phases of real-valued four-band Hamiltonians, displaying a double Hopf invariant when the bands are partitioned in two two-band subspaces~\cite{Lim2023_realhopf}, a Pontryagin index if the bands partition into a three-band and a one-band subspace~\cite{3plus1}, or two winding numbers relating to isoclinic rotations when all bands are nondegenerate, referred to as so-called generalized multigap flag phases~\cite{bouhon2020geometric, bouhon2022multigap,3plus1} (for more details, see SM~\cite{SI}).

It is in this field of interplay of geometrical insights, non-Abelian multigap notions in three-dimensional systems, and nonlinear optical responses, that we find our results. Namely, in this work, we discover quantized photovoltaic integrated shift effect in three-dimensional non-Abelian topological insulators enjoying a reality condition. In particular, we show that (i) the shift response in these systems is purely governed by the virtual transitions in the topological multiband subspaces, which (ii) we identify with the nontriviality of the torsion tensor introduced in the Riemannian-geometric reformulation of the optical responses~\cite{Ahn_2021rio}. Correspondingly, we retrieve its quantization from the real Chern-Simons forms (iii), directly related to the aforementioned Hopf indices. We numerically validate these findings in the minimal models capturing the relevant non-Abelian topologies (iv). In this context, we identify (v) the {\it quantized changes} of the shift response induced by the virtual transitions over topological phase transitions (TPTs). Finally, we discuss the novel $quantized$ integrated shift effect in the context of candidate materials.

\begin{figure}[t!]
\centering
  \includegraphics[width=1.0\columnwidth]{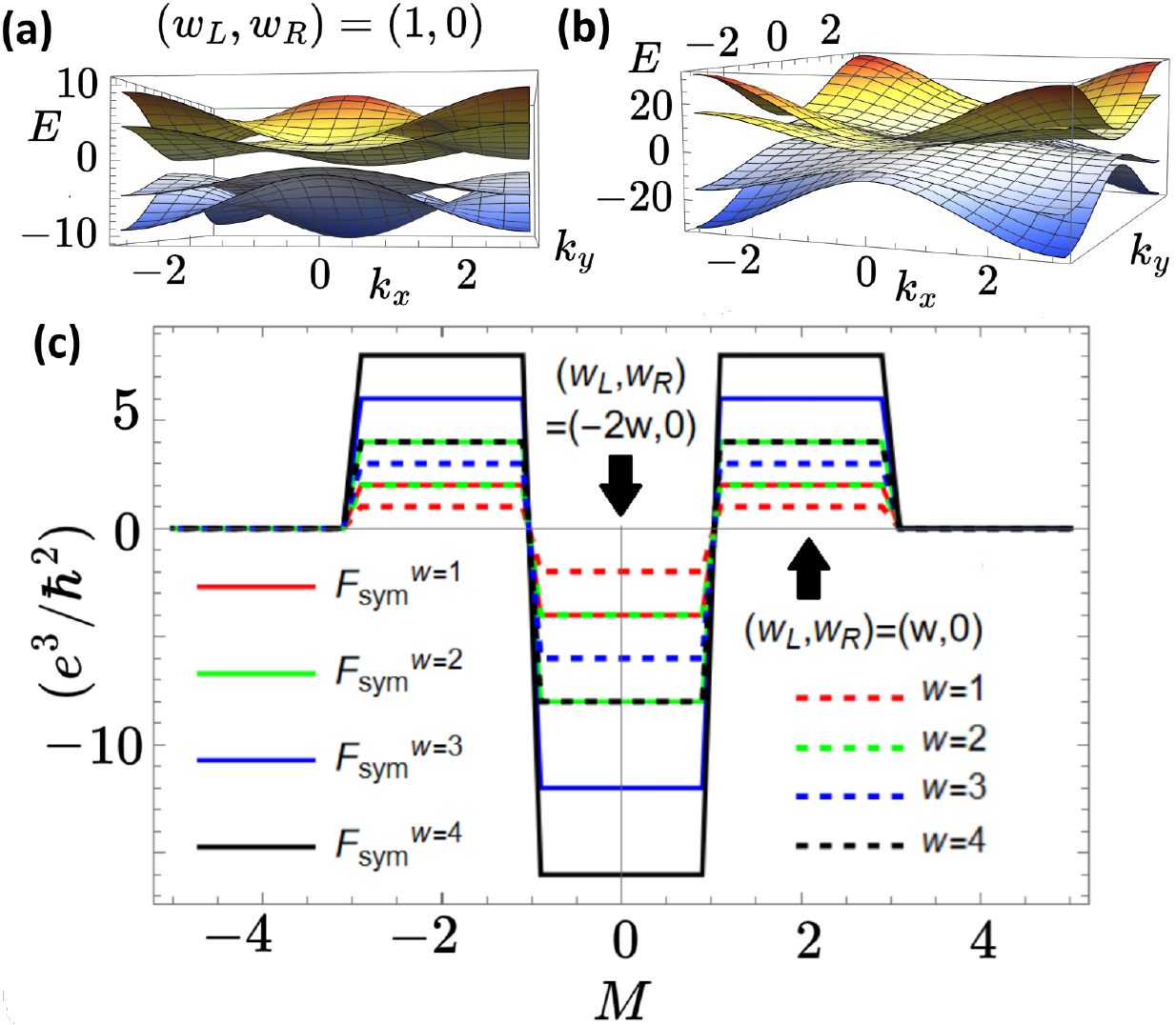}
  \caption{ \textbf{(a)} Topological band structure of the non-Abelian multigap flag phase for {$k_z = \pi/2$}.
  \textbf{(b)} Band structure at the critical topological mass $M = 3$ of the flag phase with a node emerging at the BZ center ($\kv = \textbf{0}$) at the critical point corresponding to the TPT.
  \textbf{(c)} The quantized shift effect represented by $F_{\text{sym}}$, changing sharply with the topological mass (\textit{solid}), across the values $|M| = 3/2$, ${|M| = 3}$. Correspondingly, the bulk topological invariant changes (\textit{dashed}), and trivializes further, $(w_L,w_R) = (-2w,0) \rightarrow (w,0) \rightarrow (0,0)$, for any chosen winding number $w \in \mathbb{Z}$.}
\label{fig:flag}
\end{figure}

\sect{Nonlinear optical shift responses}
The nonlinear second-order optical shift responses central to this work are given by bulk dc photovoltaic currents~\cite{1982JETP...56..359B},
\beq{}
    j_c(0) = 2\sum_{a,b=x,y,z} \sigma^{cab}_{\text{shift}}(\om) \mathcal{E}_a(\om) \mathcal{E}_b(-\om), 
\eeq
on coupling a 3D material to ac 
electric field with components of magnitude $\mathcal{E}_a(\om)$, where $\om$ is the frequency of incident electromagnetic radiation. The second-order shift photoconductivity $\sigma^{cab}_{\text{shift}}(\om)$, which can be decomposed in terms of linear $\sigma^{cab}_{\text{shift,L}}(\om)$ and circular $\sigma^{cab}_{\text{shift,C}}(\om)$ contributions picked by the corresponding light polarizations (${\sigma^{cab}_{\text{shift}}(\om) = \sigma^{cab}_{\text{shift,L}}(\om) + \ii\sigma^{cab}_{\text{shift,C}}(\om)}$)~\cite{Ahn2019}  was formulated in terms of the momentum-space derivatives of the QGT,
encoded by a Hermitian connection $C^{mn}_{abc}$~\cite{Ahn_2021rio} (see also SM~\cite{SI}). More explicitly, $C^{mn}_{bca} = -\ii r^b_{nm} r^a_{mn} R^{c,a}_{mn}$~\cite{doi:10.1126/sciadv.1501524, Ahn_2021rio}, in terms of the transition dipole moments $r^a_{nm} = \ii(1-\delta_{mn})A^a_{nm}$, given by the non-Abelian Berry connection $A^a_{nm} = \bra{u_{n,\kv}}\ket{\partial_{k_a} u_{m,\kv}}$, with $\ket{\psi_{n,\kv}} = e^{\ii\kv \cdot \rv} \ket{u_{n,\kv}}$, the Bloch states. Here, $R^{c,a}_{mn} \equiv A^c_{mm} - A^c_{nn} + \ii\partial_{k_c} \text{log}~r^a_{mn}$ is the ``shift vector", associated with the name of the effect. Intuitively, $R^{c,a}_{mn}$ reflects a positional shift of an electron upon optical transition~\cite{Sipe1,Sipe2}. In terms of $C^{mn}_{abc}$, the photoconductivity reads~\cite{Ahn_2021rio},
\begin{small}
\beq{}
    \sigma^{cab}_{\text{shift}}(\om) = \frac{\pi e^3}{2\hbar^2} \sum_{m,n} \int_{\text{BZ}} \frac{\dd^3 \kv}{(2\pi)^3} \delta(\om - \om_{mn}) f_{nm} \ii\left(C^{mn}_{acb}-C^{nm}_{bca}\right).
\eeq
\end{small}
In the above, an integration of the Hermitian connection over the entire Brillouin zone (BZ) is assumed, $f_{nm} \equiv f_{n\kv} - f_{m\kv}$ are the differences in the Fermi occupation factors of correspondingly occupied ($f_{n\kv}$) and unoccupied ($f_{m\kv}$) bands $n,m$, which are split by energies $\hbar\om_{mn}$. Importantly, the key ingredient of the response, $C^{mn}_{abc}$, defines the torsion tensor $T^{mn}_{abc}$, a~quantity central to this work,
\beq{}
    T^{mn}_{abc} \equiv C^{mn}_{abc} - C^{mn}_{acb}.
\eeq
Following Ref.~\cite{Ahn_2021rio}, we notice that physically the torsion tensor is associated with virtual transitions between bands $n,m$, intermediately through $p$ (for reference, see Fig.~\ref{fig:schematics} and SM~\cite{SI}). The question is, if such transitions can result in a topological quantized response in certain scenarios. In the following, we provide a positive answer. 

\sect{Results}
The central result of this work is to recognize that the integrated $circular$ shift photoconductivity relevant for the net bulk photovoltaic responses~\cite{alexandradinata2022topological}, can be expressed in terms of the torsion tensor (see SM~\cite{SI} for a full derivation),
\beq{eq:torsionPhoto}
\begin{split}
    \int^{\om_{\text{max}}}_0 \dd \om~ [\sigma^{abc}_{\text{shift,C}}(\om) + \sigma^{bca}_{\text{shift,C}}(\om) + \sigma^{cab}_{\text{shift,C}}(\om)] \\= -\frac{e^3}{8\pi^2 \hbar^2} \int_{\text{BZ}}  \dd^3 \kv \sum_{m,n} f_{nm} (T^{mn}_{abc} + T^{mn}_{bca} + T^{mn}_{cab}),
\end{split}
\eeq
where a frequency cutoff ${\om_{\text{max}} = \text{max} \{ (E_{m\kv} - E_{n\kv})/\hbar} \}$ is imposed for all the bands $m,n$, between which the photoexcitations are induced. Alternatively, one could set $\om_{\text{max}} \rightarrow \infty$, retrieving a net response quantized under an infinite upper limit, under which other trivial bands do not contribute in total (see SM~\cite{SI}). Notably, similar limits apply to the quantized circular dichroism that can be analogously expressed in terms of Berry curvature elements enjoying another quantization due to Chern numbers in two-dimensional systems~\cite{doi:10.1126/sciadv.1701207,Asteria_2019}. Crucially, we recognize that it is naturally possible to induce torsion in any four real Bloch bands close to the Fermi level $\{\ket{u_{1,\kv}},\ket{u_{2,\kv}},\ket{u_{3,\kv}},\ket{u_{4,\kv}} \}$, which we retrieve in the multigap topological insulators considered below. Here, two bands are occupied and two are unoccupied. Correspondingly, we provide schematics for the virtual transitions associated with the torsion in Fig.~\ref{fig:schematics}. Furthermore, on additionally considering the zero-temperature limit; $f_{nm} = 1$, we retrieve (see SM~\cite{SI}):
\beq{eq:rCS}
\begin{split}
    F_{\text{sym}} \equiv -\ii \int^{\om_{\text{max}}}_0 \dd \om~ [\sigma^{xyz}_{\text{shift}}(\om) + \sigma^{yzx}_{\text{shift}}(\om) + \sigma^{zxy}_{\text{shift}}(\om)] \\= \frac{e^3}{4\pi^2 \hbar^2} \int_{\text{BZ}} \dd^3 \kv~
    (\vec{a}^v \cdot \text{\vec{Eu}}^v + \vec{a}^c \cdot \text{\vec{Eu}}^c),
\end{split} 
\eeq
as the linear part vanishes by the reality condition, ${\sigma^{cab}_{\text{shift,L}}(\om) = 0}$~\cite{SI, Ahn_2021rio, jankowski2023optical}. Following the notion of the ``figure of merit"~of Ref.~\cite{alexandradinata2022topological}, here we spectrally integrate the shift photoconductivities to obtain a symmetrized analog, $F_{\text{sym}}$. Furthermore, we define Euler connections and curvatures~\cite{bouhon2019nonabelian} in terms of the pairs of conduction and valence bands ${\vec{a}^{v/c} = \bra{u_{2/4,\kv}} \nabla_\kv\ket{u_{1/3,\kv}}}$, ${\text{\vec{Eu}}^{v/c} = \nabla_\kv \cross \vec{a}^{v/c}}$~\cite{Lim2023_realhopf}. 
As shown in SM~\cite{SI}, the integrand can be written as the sum of Chern-Simons {3-forms}, given by non-Abelian Berry connection matrices $A^a_{nm}$ as $\text{Tr}[{\varepsilon_{abc}(A^a_{nm} \partial_{k_b} A^c_{mn} + \frac{2}{3} A^a_{nm} A^b_{mp} A^c_{pn})]}$, constrained under the reality condition, which provides a quantization upon integrating over the compact $\text{BZ}$ 3-torus. Thus, we obtain a quantized second-order photovoltaic shift effect on coupling to circularly polarized light, purely due to the virtual transitions in the real non-Abelian topological bands, i.e., bands satisfying the reality condition, and hosting topological invariants that can be captured only with non-Abelian Berry connections and curvatures~\cite{bouhon2019nonabelian,Jiang2021}. We thus identify a second-order, yet integrated, shift current cousin of the QCPGE manifested by the second-order injection currents in the Weyl semimetals~\cite{de_Juan_2017, Ma_2017}.

\begin{figure}[t!]
\centering
  \includegraphics[width=1.0\columnwidth]{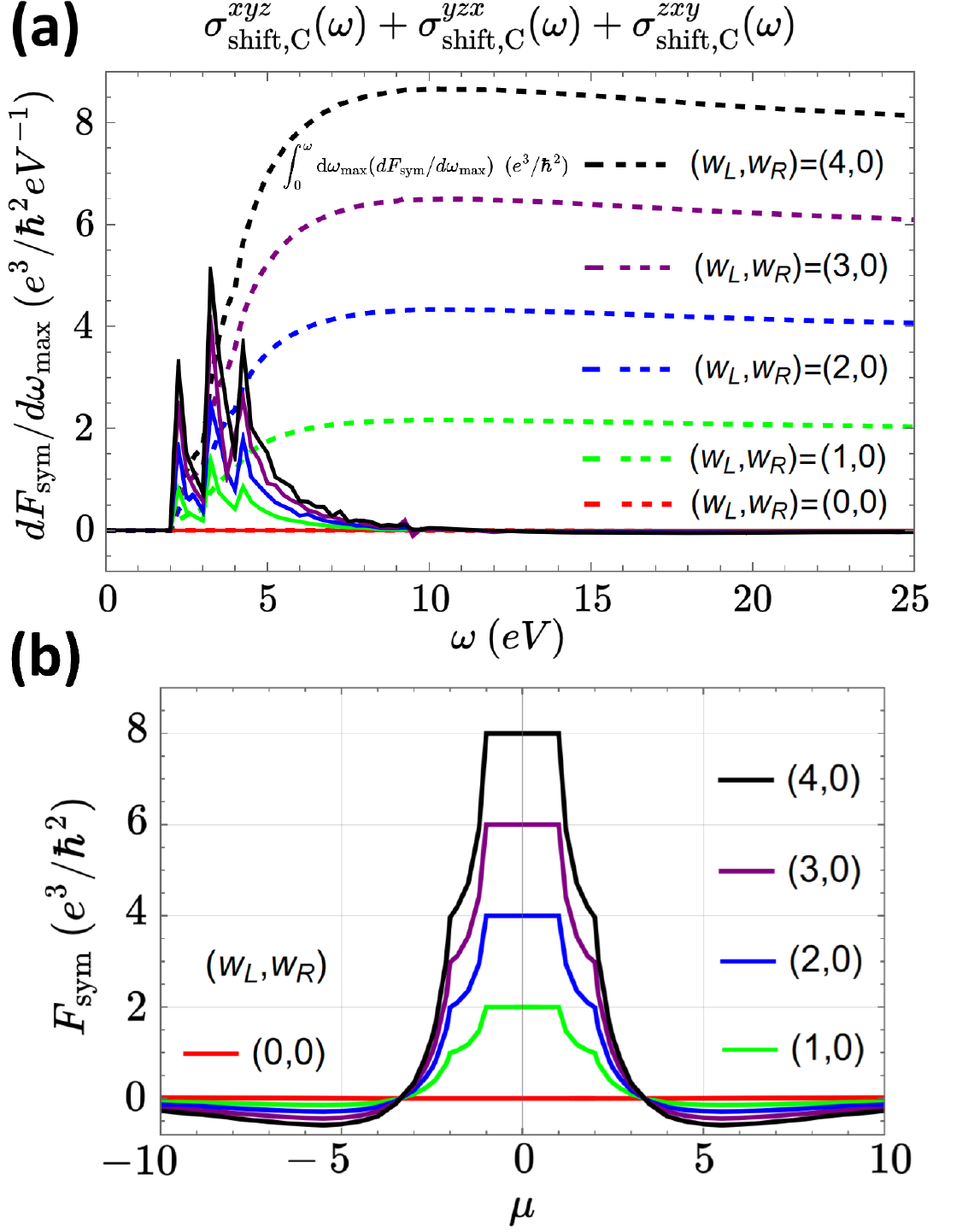}
  \caption{\textbf{(a)} Frequency-resolved shift photoconductivities constituting the~integrand of $F_{\text{sym}}$, in multigap flag phases with topological invariants $(w_L, w_R)$. The integrand of $F_{\text{sym}}$ as a function of the frequency $\om$ (\textit{solid}), and the integral of the integrand up to $\om_{\text{max}} = \om$ (\textit{dashed}) were plotted. The quantization of $F_{\text{sym}}$ is preserved as long as $\om_{\text{max}}$ does not admit resonant photoexcitations between partly occupied bands. \textbf{(b)}~$F_{\text{sym}}$ as a function of the chemical potential $\mu$. We observe that the quantization is preserved manifestly, only when $\mu$ falls into a gap between the topological bands.}
\label{fig:freq}
\end{figure}

We recognize that the quantized response holds for a broad class of non-Abelian topological insulators satisfying the reality condition imposed by $\mathcal{P}\mathcal{T}$ symmetry. In particular, this includes the recently introduced multigap topological phases classified by real flag manifolds and isoclinic rotation winding numbers $w_L, w_R \in \mathbb{Z}$ under homotopy, $\pi_3[\mathsf{Fl}_{1,1,1,1}(\mathbb{R})] \cong \mathbb{Z}^2$~\cite{3plus1}. Here, all four bands are isolated by topological gaps, all admitting topological edge modes induced by the bulk invariants and $\pi$-Zak phases~\cite{3plus1}. The invariants $w_L, w_R$ can be expressed in terms of the non-Abelian Berry connections defining Chern-Simons 3-forms of Eq.~\eqref{eq:rCS}.  See also~SM~\cite{SI} for more details on the classification of flag manifolds and multigap non-Abelian topological insulators. According to our general theory, we obtain:
\beq{eq:Fflag}
\begin{split}
    F_{\text{sym}} = \frac{2e^3}{\hbar^2} (w_L + w_R).
\end{split} 
\eeq

We reiterate that the topological quantization of the integrated shift photoconductivities holds in a broad class of topological phases 
admitting nontrivial torsion tensor, see Fig.~\ref{fig:schematics}. As another example we mention real Hopf insulators~\cite{Lim2023_realhopf} (RHIs), i.e.,~topological insulators which also require two occupied, and unoccupied Bloch bands, but enjoy band degeneracies within two-band subspaces. These are classified by a double Hopf fibration with homotopy, ${\pi_3[S^2 \cross S^2] \cong ~\mathbb{Z} \oplus \mathbb{Z}~\equiv~\mathbb{Z}^2}$~\cite{Lim2023_realhopf} (see SM~\cite{SI} for details). Here, the integrated 3-form introduced in Eq.~\eqref{eq:rCS} can be recast in terms of two real Hopf invariants, $\chi_z, \chi_w \in \mathbb{Z}$:
\beq{eq:FHopf}
\begin{split}
    F_{\text{sym}} = \frac{2e^3}{\hbar^2} (\chi_z + \chi_w).
\end{split} 
\eeq

Yet another example pertains to the $3\oplus1$ band partitioning of the four  bands. In this case the system is classified by a single topological index, being the Pontryagin index $Q \in \mathbb{Z}$, defined by $\pi_3[S^3] \cong \mathbb{Z}$~\cite{3plus1}. As before, a perfectly quantized response (${F_{\text{sym}} = \frac{2e^3}{\hbar^2}} Q$) could be achieved~\cite{SI}. 
This completes the glossary of the quantized integrated shift responses due to different types of four-band topologies. Most importantly, we underpin the generality of the effect by showing the stability upon adding extra bands. Namely, we show in the SM~\cite{SI} that the quantization of $F_{\text{sym}}$ persists as long as the gaps above and below the four-band subspace are preserved and $\om_{\text{max}}$ targets the energy window between these gaps.

\sect{Model realizations}
We can illustrate the above ideas in the context of multigap topological insulator models. For details on general classes of models capturing the considered band topologies, we furthermore refer to SM~\cite{SI}. We employ models with tunable topological mass $M$. In particular, we choose ${H(\kv) = V_R V^{\text{T}}_L~\text{diag}[-2,-1,1,2]V_L V^{\text{T}}_R}$ with,
\beq{}
\begin{split}
    V_{L/R} = \sin w_{L/R} k_x~\Gamma_{11} + \ii \sin k_y~\Gamma_{23} - \sin k_z~\Gamma_{31} +\\+ (M - \cos w_{L/R} k_x - \cos k_y - \cos k_z)~\Gamma_{03},
\end{split}
\eeq
in terms of $4 \times 4$ Dirac matrices $\Gamma_{ij} = \sigma_i~\otimes~\sigma_j$ defined as Kronecker products of identity and Pauli matrices $(\sigma_0, \sigma_1, \sigma_2, \sigma_3) \equiv (\mathbf{1}_2, \sigma_x, \sigma_y, \sigma_z)$, and the winding numbers $w_L, w_R$, equal to the topological invariants in the regime ${-3/2 < |M| < 3}$. For $|M| > 3$, the Hamiltonians are topologically trivial when $F_{\text{sym}}$ vanishes. For $|M| < 3/2$, the invariants can be doubled in magnitude, see Fig.~\ref{fig:flag}. In Fig.~\ref{fig:flag}, we correspondingly show sharp quantized jumps in the integrated shift responses $F_{\text{sym}}$, coinciding with the values of topological masses corresponding to TPTs. At the TPT points, the systems become semimetallic at gap closings, and $F_{\text{sym}}$ loses quantization across jumps. Notably, the shift photoconductivities in topological semimetals, including $\mathcal{PT}$-symmetric models, were extensively studied in Refs.~\cite{AhnPRX, PhysRevB.107.155434, PhysRevB.107.L201112, PhysRevLett.131.116603}.

In Fig.~\ref{fig:freq}, we further plot the frequency-resolved integrand of $F_{\text{sym}}$, demonstrating how the quantization is retrieved on integration over the frequency domain. We note that there are three peaks yielding dominant contributions to the integer-valued quantization of $F_{\text{sym}}$ in the studied models, which correspond to transitions: $\ket{u_{2,\kv}} \rightarrow \ket{u_{3,\kv}}$, $\ket{u_{1/2,\kv}} \rightarrow \ket{u_{3/4,\kv}}$, $\ket{u_{1,\kv}} \rightarrow \ket{u_{4,\kv}}$. We moreover show the robustness of the quantization to changing the chemical potential, finding that any doping away from the gap removes the quantization, as expected.

\sect{Discussion}
Having introduced the topological quantized phenomenology, we address the possibilities of observing such physical effects and how our results galvanize new areas of investigation. First, we stress that our bulk quantization conditions of the integrated shift responses are distinct from other interesting conditions, such as the quantization of the individual shift vectors~\cite{alexandradinata2022topological}. Moreover, unlike QCPGE, the introduced effect relies on performing a spectral summation, similarly to the quantized circular dichroism found in Chern insulators~\cite{doi:10.1126/sciadv.1701207, Asteria_2019}. Interestingly, the effect is manifestly dissipationless, i.e., the response is nondissipative, as the photocurrent components are orthogonal to the electric fields carried by the incident circularly polarized light. Importantly, we stress that the shift effect considered in this work is purely associated with photoexcitations, rather than with the recombination or intraband parts of the shift current~\cite{1982JETP...56..359B,zhu2024anomalousshiftopticalvorticity}. Experimentally, such excitation-dominated responses can be measured in photoconductivity experiments with subpicosecond resolution~\cite{Sotome2019, zhu2024anomalousshiftopticalvorticity}. Furthermore, one could wonder about contributions to the shift currents by the 2D surface modes corresponding to the nontrivial non-Abelian 3D topological bulk. Notably, the surface bands are unlikely to contribute to $F_{\text{sym}}$, as it by construction involves combinations of all three spatial dimensions in each constituent photoconductivity term. Finally, an interesting future direction entails the robustness of the quantized effect under disorder or correlations, as the quantization of QCPGE breaks down under such effects~\cite{Avdoshkin_2020, Wu_2024}. 

We identify $\mathcal{PT}$-symmetric 3D topological phases, among the candidate materials to realize the effect. Such phases include CuMnAs, a system modeled with a four-band low-energy theory, already known to support chiral gyration photocurrents, that is~$j_x$ induced by Hermitian connection elements $C^{mn}_{xxy}, C^{mn}_{yxy}$, under circularly polarized light~\cite{PhysRevX.11.011001, PhysRevB.104.024416}, or phases obtainable from anti-ferromagnetic MnBi$_{2n}$Te$_{3n+1}$ heterostructures~\cite{otrokov2019prediction, mong2010_afm,Jo2020_afm}, which were recently related to nontrivial quantum geometry~\cite{Gao_2023}. In particular, a measurable quantized jump in $F_{\text{sym}}$ is expected across TPT, if the considered multigap topologies were present. We stress that the quantization of $F_{\text{sym}}$ is protected by the gaps from additional bands,
as we also further discuss in~\cite{SI}. Importantly, the quantization is robust for the topological bands that do not hybridize with any other bands, i.e., have vanishing electric dipole matrix elements for transitions to the other bands, as can be ensured by the orbital symmetries and selection rules, see SM~\cite{SI}. Under generic hybridizations, the quantization of both the invariant and $F_{\text{sym}}$ could be potentially lost, as we further address in SM~\cite{SI}. Crucially, to obtain the nontrivial second-order shift effect, the material needs to be non-centrosymmetric~\cite{Sipe1}, i.e., break $\mathcal{P}$ symmetry. Hence, to preserve $\mathcal{PT}$, it also necessarily needs to break $\mathcal{T}$ symmetry, i.e., be magnetic. Indeed, $\mathcal{P}$ symmetry is admitted only if $w_L + w_R = 0$, or $\chi_w + \chi_z = 0$ for RHI, as shown in Ref.~\cite{Lim2023_realhopf}, which yields no shift response, consistent with Eqs.~\eqref{eq:Fflag},~\eqref{eq:FHopf}.

We finally note that recently it was suggested that virtual transitions can constitute a significant contribution to shift currents in the multiband systems, in particular, layered moir\'e materials~\cite{chen2024enhancing}, and that there might exist interesting connections to the band topology. Here, we precisely retrieve such a connection, manifestly captured by a quantization encoded in the virtual multiband transitions. Additionally, we show that indeed, the changes in the band topology across TPTs directly result in the quantized jumps of the shift responses present in non-Abelian multiband topological insulators. Relating the above results and multigap phases to moir\'e platforms, which can be expanded to three-dimensional setups~\cite{wang2024threedimensionalmoirecrystal}, therefore generally entails for a promising future pursuit.

\sect{Summary and Outlook}
We identify a new \textit{quantized} photovoltaic response due to (circular) shift photoconductivities, which is manifested by second-order currents. We analytically and numerically retrieve this response in topological phases with flag manifold, real Hopf, and Pontryagin-indexed, multigap topologies, captured by minimal four Bloch band models under the reality conditions. In particular, we show that the integrated shift effect, purely contributed by the virtual interband transitions, can indicate topological phase transitions in non-Abelian topological phases. Hence, we find another unique topological quantized dc response, beyond QAH effect in Chern insulators and QCPGE in Weyl semimetals, providing a smoking-gun criterion for non-Abelian multigap topological phases.\\

\begin{acknowledgements}
    W.J.J.~acknowledges funding from the Rod Smallwood Studentship at Trinity College, Cambridge. R.-J.S. acknowledges funding from a New Investigator Award, EPSRC Grant No. EP/W00187X/1, a EPSRC ERC underwrite Grant No.  EP/X025829/1, and a Royal Society exchange Grant No. IES/R1/221060, as well as Trinity College, Cambridge. We cordially thank Zory Davoyan, Adrien Bouhon, and F. Nur \"Unal for numerous discussions concerning non-Abelian and multigap topological phases. We thank Aris Alexandradinata for discussions about shift photocurrents. R.-J.~S thanks Joel E. Moore for discussions. \\

 \textit{Note added:} While completing this manuscript, we became aware of the work \cite{chen2024enhancing} on enhancing shift currents via multiband transitions, although in the 2D context of moir\'e materials. Here, beyond retrieving an unprecedented $\textit{quantized}$ shift response due to the virtual transitions in 3D systems, we explictly demonstrate shift responses for probing TPTs in 3D multiband topological materials.
\end{acknowledgements}

\bibliography{references}

\newpage

\end{document}


\title{SUPPLEMENTAL MATERIAL \\ Quantized Integrated Shift Effect in Multigap Topological Phases}

\author{Wojciech J. Jankowski}
\email{wjj25@cam.ac.uk}
\affiliation{TCM Group, Cavendish Laboratory, Department of Physics, J J Thomson Avenue, Cambridge CB3 0HE, United Kingdom}
 
\author{Robert-Jan Slager}
\email{rjs269@cam.ac.uk}
\affiliation{TCM Group, Cavendish Laboratory, Department of Physics, J J Thomson Avenue, Cambridge CB3 0HE, United Kingdom}

\date{\today}

\maketitle

\section{Basic quantum-geometric definitions}\label{app::A}

In this section, for completeness, we define the relevant quantum-geometric objects within the single-particle picture of Bloch states, as originally introduced in Ref.~\cite{provost1980riemannian} and extended towards optical responses in Ref.~\cite{Ahn_2021rio}. Following Ref.~\cite{Ahn_2021rio}, we start with the quantum-geometric tensor (QGT) for bands $m,n$, as mentioned in the main text,
\beq{}
    Q^{mn}_{ab} \equiv r^a_{nm}r^b_{mn}.
\eeq{}
Here, the transition dipole moment vector, $\rv_{mn}(\kv) = \bra{\psi_{m,\kv}} \hat{\rv} \ket{\psi_{n,\kv}}$, provides a probability amplitude for an electric dipole (E1) transition between Bloch states $\ket{\psi_{n,\kv}}$ and $\ket{\psi_{m,\kv}}$. More abstractly, $\rv_{mn}$ defines a tangent vector on the manifold of quantum states~\cite{Ahn_2021rio}, with components
\beq{}
    \rv^a_{mn}(\kv) = \ii(1-\delta_{mn}) A^{a}_{nm}(\kv)
\eec
where $A^{a}_{nm}(\kv) \equiv \bra{u_{n,\kv}}  \ket{\partial_{a} u_{m,\kv}}$ is the non-Abelian Berry connection, and we employ a simplified notation $\partial_a \equiv \partial_{k_a}$.

The QGT, known otherwise as Hermitian metric, decomposes as
\beq{}
    Q^{mn}_{ab} = g^{mn}_{ab} - \frac{\ii}{2}F^{mn}_{ab},
\eeq{}
and equips the vector bundle of Bloch states~\cite{bouhon2023quantum} over a Brillouin zone (BZ), $\text{BZ} \cong T^d$, where $T^d$ is a $d$-dimensional torus, with a Riemannian structure through its real part $g^{mn}_{ab} = \mathfrak{Re}~Q^{mn}_{ab}$, the quantum metric. In particular $g^{mn}_{ab}$ defines a Fubini-Study metric through $g_{ab}^n = \sum^\text{unocc}_{m} g^{mn}_{ab}$,
\beq{}
    ds^2_n \equiv 1 - |\bra{u_{n,\kv}}\ket{u_{n,\kv+\delta \kv}}|^2 = g_{ab}^n dk_a dk_b, 
\eeq
for any band $n$. The symplectic part $F^{mn}_{ab}$ yields the Abelian Berry curvature $F^n_{ab} = \sum^\text{unocc}_{m} F^{mn}_{ab}$ on performing a summation over the unoccupied bands.

Furthermore, for higher order in $k$-space derivatives, the Hermitian connection can be defined~\cite{Ahn_2021rio}.
The advantage of the described representation is that it allows to easily introduce a Hermitian connection as
\beq{}
    C^{mn}_{abc} \equiv r^a_{nm}r^{c;b}_{mn},
\eeq
where we include the covariant derivative $r^{c;b}_{mn}$ of the transition dipole, defined in terms of the Berry connection as
\\
\beq{cov}
\begin{split}
    r^{c;b}_{mn} (\kv) = \nabla_{b} r^{c}_{mn} (\kv) = \partial_{b} r^{c}_{mn} (\kv) - \ii(A^{b}_{nn} (\kv) - A^{b}_{mm} (\kv)) r^{c}_{mn} (\kv).
\end{split}
\eeq
The formal definition in terms of covariant derivative can be directly recast into the shift vector definition introduced in the main text.

Furthermore, the Hermitian connection can be decomposed in terms of the metric connection $\mathfrak{Re}~C^{mn}_{abc}$ and symplectic connection $\mathfrak{Im}~C^{mn}_{abc}$ as
\beq{}
    C^{mn}_{abc} = \mathfrak{Re}~C^{mn}_{abc} + \ii \mathfrak{Im}~C^{mn}_{abc}.
\eeq
The \textit{torsion-free} part of the metric connection $\mathfrak{Re}~C^{mn}_{abc}$ reads~\cite{AhnPRX},
\beq{}
    \Gamma^{mn}_{abc} = \frac{1}{2}(\partial_b g^{mn}_{ac} + \partial_c g^{mn}_{ab} - \partial_a g^{mn}_{bc}),
\eeq
but this part crucially omits the virtual transitions~\cite{Ahn_2021rio}, essential for the optical responses of the non-Abelian topological insulators. The torsion tensor $T^{mn}_{abc} \equiv C^{mn}_{abc}-C^{mn}_{acb}$, as defined in the main text, explicitly captures the virtual transitions as~\cite{Ahn_2021rio},
\beq{}
    T^{mn}_{abc} = \ii r^a_{n m} \sum_{p \neq m, n} (r^b_{m p} r^c_{p n} - r^c_{m p} r^b_{p n}),
\eeq
on direct substition in terms of $r^a_{mn}$ transition dipole matrix elements. 
Explicitly, the torsion tensor elements combine the E1 transition amplitudes given by the transition dipole matrix elements $r^a_{mn}$ for (i) starting from a occupied band $n$, (ii) virtually exciting an electron through band $p$, and (iii) finally ending up photoexcited in initially unoccupied band $m$.

We note that in particular, if the Bloch states can be chosen in the real gauge, i.e. $\ket{u_{n,\kv}}$ become real vectors, under the so-called reality condition (see Sec.~\ref{app::B}), $C_{abc} = \mathfrak{Re}~C_{abc}$, as follows from a direct substitution of Bloch states to the transition dipole moment matrix elements. 

\section{Non-Abelian three-dimensional topologies under the reality condition}\label{app::B}

In this section, we first clarify the constraints due to the reality condition imposed by the $\mathcal{P}\mathcal{T}$ symmetry imposed on three-dimensional non-Abelian topological phases. Regarding the latter, we provide simple models and elaborate on the topologies of (i) multigap flag phases, (ii) the real Hopf insulator, (iii) Pontryagin-indexed insulator, all of which can be generated and characterized within the Pl\"ucker formalism framework~\cite{bouhon2020geometric,bouhon2023quantum}. Furthermore, following Ref.~\cite{bouhon2020geometric}, the momentum-space Bloch Hamiltonians of multigap topological phases can be Fourier transformed to their real-space representation. Consequently, these Hamiltonians can be expressed in terms of Wannier orbital basis that encodes the real-space hopping amplitudes.

\subsection{Reality condition}

The action of the $\mathcal{P}\mathcal{T}$ symmetry (here, we consider spinless $(\mathcal{P}\mathcal{T})^2 =+1$ symmetry) ensures that the Bloch Hamiltonian of a three-dimensional solid treated within an independent-particle approximation is real, $H(\kv) = H^{*}(\kv)$, and there exists a gauge in which the associated eigenstates labelled with quasimomenta $\kv = (k_x,k_y,k_z)^{\text{T}}$ are also real vectors. In particular, this follows from the symmetry transformations of the Hamiltonian~\cite{bouhon2020geometric}:
\beq{}
    (\mathcal{PT}) H(k_x,k_y,k_z) (\mathcal{PT})^{-1} = H^*(k_x,k_y,k_z) = H(k_x,k_y,k_z),  
\eeq
which is related to the fact that the individual symmetries of the composed symmetry act as:
\beq{}
    \mathcal{T} H(k_x,k_y,k_z) \mathcal{T}^{-1} = H^*(-k_x,-k_y,-k_z),  
\eeq
\beq{}
    \mathcal{P} H(k_x,k_y,k_z) \mathcal{P}^{-1} = H(-k_x,-k_y,-k_z).  
\eeq
Hence, the Bloch Hamiltonian, $H(k_x,k_y,k_z) = H^{*}(k_x,k_y,k_z)$, can be represented as a real matrix~\cite{bouhon2020geometric}. Combined with the Hermiticity of the Hamiltonian, the Hamiltonian indeed reduces to a real $symmetric$ matrix, therefore it enjoys a well-known property that its eigenvectors can be taken real, $\ket{u_{n,\kv}} \in \mathbb{R}^N$~\cite{bouhon2020geometric}.

Therefore, we note that under such reality condition imposed by the $\mathcal{P}\mathcal{T}$ symmetry,  
\beq{}
    \vec{A}_{nn}(\kv) = \ii\bra{u_{n,\kv}} \ket{\nabla_\kv u_{n,\kv}} = \ii\nabla_\kv \bra{u_{n,\kv}}\ket{u_{n,\kv}} - \ii\bra{\nabla_\kv u_{n,\kv}}  \ket{u_{n,\kv}} = - \ii\bra{ u_{n,\kv}}\ket{  
 \nabla_\kv u_{n,\kv}} = -\vec{A}_{nn}(\kv),
\eeq
using (i) a product rule, (ii) normalization condition imposed on the eigenvectors, and (iii) complex conjugation of a real inner product. Hence, $\vec{A}_{nn}(\kv) = 0$, i.e. the diagonal elements of the non-Abelian Berry connection, as well as its trace, vanish identically; as a result of imposing the reality condition on the Hamiltonian.

\subsection{Multigap flag phases}

Here, we provide details on non-Abelian real topology of multigap flag phases, including the construction of relevant models encoded in corresponding Bloch Hamiltonians. As introduced in Ref.~\cite{3plus1}, the multigap flag phases with four isolated bands are classified by the homotopy of flag varieties, $\widetilde{\mathsf{Fl}}_{p_1,\ldots,p_k}(\mathbb{R}) \equiv \mathsf{SO}(\sum^k_{i=1} p_i)/[\mathsf{SO}(p_1) \times \ldots \times \mathsf{SO}(p_k)]$, where $p_1, \ldots, p_k \in \mathbb{N}$ are numbers of bands in the gapped subbundles. Here, $p_1 = p_2 = p_3 = p_4 = 1$, with:
\\
\beq{}
    \widetilde{\mathsf{Fl}}_{1,1,1,1}(\mathbb{R}) = \mathsf{SO}(4)\cong \frac{S^3 \times S^3}{\mathbb{Z}_2}.
\eeq
\\
Namely, we have $\pi_3[\widetilde{\mathsf{Fl}}_{1,1,1,1}(\mathbb{R})] \cong \pi_3[\mathsf{SO}(4)] \cong \mathbb{Z} \oplus \mathbb{Z} = \mathbb{Z}^2$.

To interpret the meaning of the $\mathbb{Z}^2$ invariant, we recall the well-known isomorphism,
\beq
    S\mathsf{SO}(4) \cong \frac{S^3_L \times S^3_R}{\mathbb{Z}_2},
\eeq
which on recognizing that the $\mathbb{Z}_2$ quotient does not affect $\pi_n ~\text{for}~ n \geq 2$, allows to interpret the invariant as a pair of windings on two copies of 3-spheres~\cite{3plus1},
\beq{}
\pi_3[S^3_L \times S^3_R] \cong \pi_3[S^3_L] \oplus \pi_3[S^3_R] \cong \mathbb{Z} \oplus \mathbb{Z}.
\eeq
We recognize that the nontrivial $S^3 \rightarrow S^3$ maps corresponding to the nontrivial elements of $\pi_3[S^3_{L/R}] \cong \mathbb{Z}$ can be interpreted as 
winding numbers $w_{L}, w_{R}$. 

Correspondingly, we construct the flag Hamiltonian by taking:
\begin{equation}
    H(\textbf{k}) = V(\textbf{k})~\text{diag} [ E_1, E_2, E_3, E_4 ]~V(\textbf{k})^{\text{T}},
\end{equation}
with $E_1 < E_2 < E_3 < E_4$ introducing three gaps, and $V = V_R V^{\text{T}}_L$ which can be factored in terms of two matrices introducing the isoclinic windings. These can be constructed as
\begin{equation}
V_L=\left(\begin{array}{rrrr}l_0 & -l_3 & l_2 & -l_1 \\ l_3 & l_0 & -l_1 & -l_2 \\ -l_2 & l_1 & l_0 & -l_3 \\ l_1 & l_2 & l_3 & l_0\end{array}\right),
\end{equation}
\\
\begin{equation}
  V_R=\left(\begin{array}{rrrr}r_0 & -r_3 & r_2 & r_1 \\ r_3 & r_0 & -r_1 & r_2 \\ -r_2 & r_1 & r_0 & r_3 \\ -r_1 & -r_2 & -r_3 & r_0\end{array}\right),  
\end{equation}
with $l_0^2 + l_1^2 + l_2^2 + l_3^2 = 1$ and $r_0^2 + r_1^2 + r_2^2 + r_3^2 = 1$.
On such factorization, the invariants of the flag phases can be directly deduced from the following formulae of the winding invariants:
\\
\begin{equation}
\begin{aligned}
&w_L = \frac{1}{2 \pi^2}\int_{\text{BZ}} \dd^3\kv~ \varepsilon_{ijpq}l^i \partial_{k_x} l^j \partial_{k_y} l^p \partial_{k_z}l^q, \\ 
&w_R = \frac{1}{2 \pi^2}\int_{\text{BZ}} \dd^3\kv~  \varepsilon_{ijpq}r^i \partial_{k_x}r^j \partial_{k_y}r^p \partial_{k_z}r^q. \\
\end{aligned}
\end{equation}
\\
In particular, we can model these by a choice
\begin{equation}
  (l_0, l_1, l_2, l_3) = (\sin{w_L k_x}, \sin{k_y},\sin{k_z}, M - \cos{w_L k_x}-\cos{k_y}-\cos{k_z})^{\text{T}},
\end{equation}
\begin{equation}
(r_0, r_1, r_2, r_3) = (\sin{w_R k_x}, \sin{k_y},\sin{k_z}, M - \cos{w_R k_x}-\cos{k_y}-\cos{k_z})^{\text{T}}.  
\end{equation}
\\
where we set the topological mass to $M=2$. Additionally, changing the parameter $M$ allows to induce topological phase transitions (TPTs), on gap closings, as shown explicitly in the main text.

Finally, we note that in terms of the non-Abelian Berry connections and Euler connections/curvatures, we can rewrite the winding invariants $(w_L, w_R)$ of four topological bands simply as~\cite{3plus1},
\beq{}
    w_L = -\frac{1}{16 \pi^2}\int_{\text{BZ}} \dd^3\kv~ (\textbf{A}_{12} + \textbf{A}_{34}) \cdot \nabla_\kv \times (\textbf{A}_{12} + \textbf{A}_{34}) = -\frac{1}{16 \pi^2}\int_{\text{BZ}} \dd^3\kv~ (\textbf{a}^c + \textbf{a}^v) \cdot (\textbf{Eu}^c + \textbf{Eu}^v),
\eeq
\beq{}
    w_R = -\frac{1}{16 \pi^2}\int_{\text{BZ}} \dd^3\kv~ (\textbf{A}_{12} - \textbf{A}_{34}) \cdot \nabla_\kv \times (\textbf{A}_{12} - \textbf{A}_{34}) = -\frac{1}{16 \pi^2}\int_{\text{BZ}} \dd^3\kv~ (\textbf{a}^c - \textbf{a}^v) \cdot (\textbf{Eu}^c - \textbf{Eu}^v), \\
\eeq
which is of the same functional form as the real Hopf invariants defined in the next subsection~\cite{Lim2023_realhopf, 3plus1}.

\subsection{Real Hopf insulator}

In the case of a real Hopf insulator, see Ref.~\cite{Lim2023_realhopf}, we consider the $2 \oplus 2$ band partitioning. In terms of homotopy classification of the topological phases encoded with real topologies~\cite{bouhon2020geometric}, the band partitioning yields a classification on an oriented real Grassmannian manifold,
\beq{}
    \widetilde{\mathsf{Gr}}_{2,4}(\mathbb{R}) = \mathsf{SO}(4)/[\mathsf{SO}(2) \cross \mathsf{SO}(2)] \cong S^2 \cross S^2.
\eeq
Therefore, the invariant can be obtained as $\pi_3[\widetilde{\mathsf{Gr}}_{2,4}(\mathbb{R})] \cong \pi_3[S^2 \cross S^2] \cong \pi_3[S^2]~\oplus~\pi_3[S^2] \cong \mathbb{Z}~ \oplus~\mathbb{Z}$, where each $\pi_3[S^2] \cong \mathbb{Z}$ defines a Hopf fibration ($S^3 \rightarrow S^2$) and an associated Hopf index. 

In particular, these invariants can be expressed by adding/subtracting products of the Euler curvature and connection elements mixing both valence and conduction bands~\cite{Lim2023_realhopf}: $\vec{a}^c \cdot \text{\vec{Eu}}^v$ and $\vec{a}^v \cdot \text{\vec{Eu}}^c$, in the integrand of Eq.~(5) of the main text. Namely, following Ref.~\cite{Lim2023_realhopf}, these Hopf invariants can be written as:

\begin{equation}
\begin{aligned} 
 \chi_z = -\frac{1}{16 \pi^2} \int_{\text{BZ}} \mathrm{a}^c \wedge \mathrm{Eu}^c+\mathrm{a}^v \wedge \mathrm{Eu}^v + \mathrm{a}^c \wedge \mathrm{Eu}^v+\mathrm{a}^v \wedge \mathrm{Eu}^c \in \mathbb{Z}, \\ 
 \chi_w = -\frac{1}{16 \pi^2} \int_{\text{BZ}}  \mathrm{a}^c \wedge \mathrm{Eu}^c+\mathrm{a}^v \wedge \mathrm{Eu}^v -\mathrm{a}^c \wedge \mathrm{Eu}^v - \mathrm{a}^v \wedge \mathrm{Eu}^c \in \mathbb{Z}.
\end{aligned}
\end{equation}

In particular, we notice that on summation, the invariants yield
\beq{}
    \chi_z + \chi_w = -\frac{1}{8 \pi^2} \int_{\text{BZ}}  \mathrm{a}^c \wedge \mathrm{Eu}^c+\mathrm{a}^v \wedge \mathrm{Eu}^v \in \mathbb{Z},
\eeq
which is the quantized number entering the integrated shift response introduced in the main text. Here, we note that while the previously introduced topological indices $w_L, w_R \in \mathbb{Z}$ can be obtained analogously to the real Hopf invariants from the non-Abelian Berry connections, those are not associated with the Hopf fibrations, which characterize the mappings of two-band subbundles to 2-spheres $S^2 \times S^2$. In other words, each winding number corresponds to a nontrivial element of $\pi_3[S^3]$, rather than $\pi_3[S^2]$. In particular, the multiply-gapped flag insulators support edge states in all three topological gaps between four topological bands~\cite{3plus1}

The models for real Hopf insulators can be obtained from a flattened Hamiltonian ensuring a correct band partitioning
\beq{}
    H^{\chi_z,\chi_w}(\kv) = R_{z,w}(\kv)~\text{\text{diag}} [-1, -1, 1, 1]~ R^{\text{T}}_{z,w}(\kv),
\eeq
where $R_{z,w}(\kv)$ encodes the double Hopf fibration inducing indices ($\chi_w,\chi_z$), as introduced and discussed in detail in Ref.~\cite{Lim2023_realhopf}.

\subsection{Pontryagin-indexed
insulator}

To define the insulator with Pontryagin index, we demand a $3 \oplus 1$ partitioning of four topological bands. From the homotopy perspective, the effective classifying space of interest becomes a oriented real Grassmannian,\\
\beq{}
    \widetilde{\mathsf{Gr}}_{1,4}(\mathbb{R}) = \mathsf{SO}(4)/\mathsf{SO}(3) \cong S^3,
\eeq
\\
as discussed in Ref.~\cite{3plus1}. The homotopy classification in three spatial dimensions then gives
\beq{}
    \pi_3 [\widetilde{\mathsf{Gr}}_{1,4}(\mathbb{R})] \cong \pi_3 [S^3] \cong \mathbb{Z},
\eeq
where indeed $\pi_3 [S^3] \cong \mathbb{Z}$ defines the Pontryagin index $Q$, which classifies homotopy-inequivalent maps: $S^3 \rightarrow S^3$~\cite{Zee2010-ZEEQFT}.

The Pontryagin index can be computed from the non-Abelian Berry connection elements as 
\\
\beq{}
    Q = \frac{1}{2\pi^2} \int_{T^3} \dd^{3}k~ \varepsilon^{ijk} A^i_{41} A^j_{42} A^k_{43}.
\eeq
\\
Equivalently~\cite{3plus1}, the invariant can be computed as the winding of the fourth band~$\vec{u}_4 \equiv \ket{u_{4,\kv}}$ as,
\\
\beq{Pont_index}
    Q = \frac{1}{2\pi^2} \int_{S^3} \dd^{3}k~
    \varepsilon_{ijpq} (\vec{u}_4)^i \partial_{k_x} (\vec{u}_4)^j \partial_{k_y} (\vec{u}_4)^p \partial_{k_z} (\vec{u}_4)^q 
\eec
\\
pushed forward to a 3-sphere $S^3$ from the BZ 3-torus ($T^3 \rightarrow S^3$). As shown in Ref.~\cite{3plus1}, to generate a concrete model, one could take a flattened Hamiltonian and a Bloch band matrix $R(\kv)$ as\\
\beq{}
    H^Q(\kv) = R(\kv)
    \begin{pmatrix}
        -1 & 0 & 0 & 0 \\
        0 & 1 & 0 & 0 \\
        0 & 0 & 1 & 0 \\
        0 & 0 & 0 & 1 \\
    \end{pmatrix} R(\kv)^{\text{T}}.
\eeq
\\
Here, the matrix $R(\kv) \in \mathsf{SO}(4)$ can be obtained by a three-angle parametrization $(\phi, \psi, \theta)$ obtained from Pl\"ucker formalism, namely~\cite{bouhon2020geometric,3plus1}:
\\
\beq{}
   \psi(\kv) = \pi~\text{max}\{|k_x|, |k_y|, |k_z|\}
\eec
\\
\beq{}
    \theta(\kv) = \cos^{-1}(k_z/\sqrt{k^2_x + k^2_y + k^2_z})
\eec
\\
\beq{}
    \phi(\kv) = Q \tan^{-1}(k_x/k_y)
\eec
\\
deduced for any $k$-point in the BZ. Correspondingly, the Bloch matrix inducing $Q \neq 0$ can be generated as
\\
\beq{}
    R(\kv) =  \text{e}^{\ii\theta\Gamma_{20}} \text{e}^{\ii\phi\Gamma_{21}}
    \text{e}^{\ii\psi\Gamma_{12}}.
\eeq
\\
In particular, the hoppings can be truncated at a finite-neighbor hopping level, yielding a local Hamiltonian, while preserving its topology, following the general Pl\"ucker embedding framework for generating models~\cite{bouhon2020geometric}. Additionally, a further class of flat-band models in Pontryagin-indexed insulators was studied in detail in Ref.~\cite{3plus1}, where further parametrizations were provided. This concludes the summary of the previously-reported possible non-Abelian four-band topologies in three spatial dimensions.
\begin{figure}
\centering
  \includegraphics[width=0.5\columnwidth]{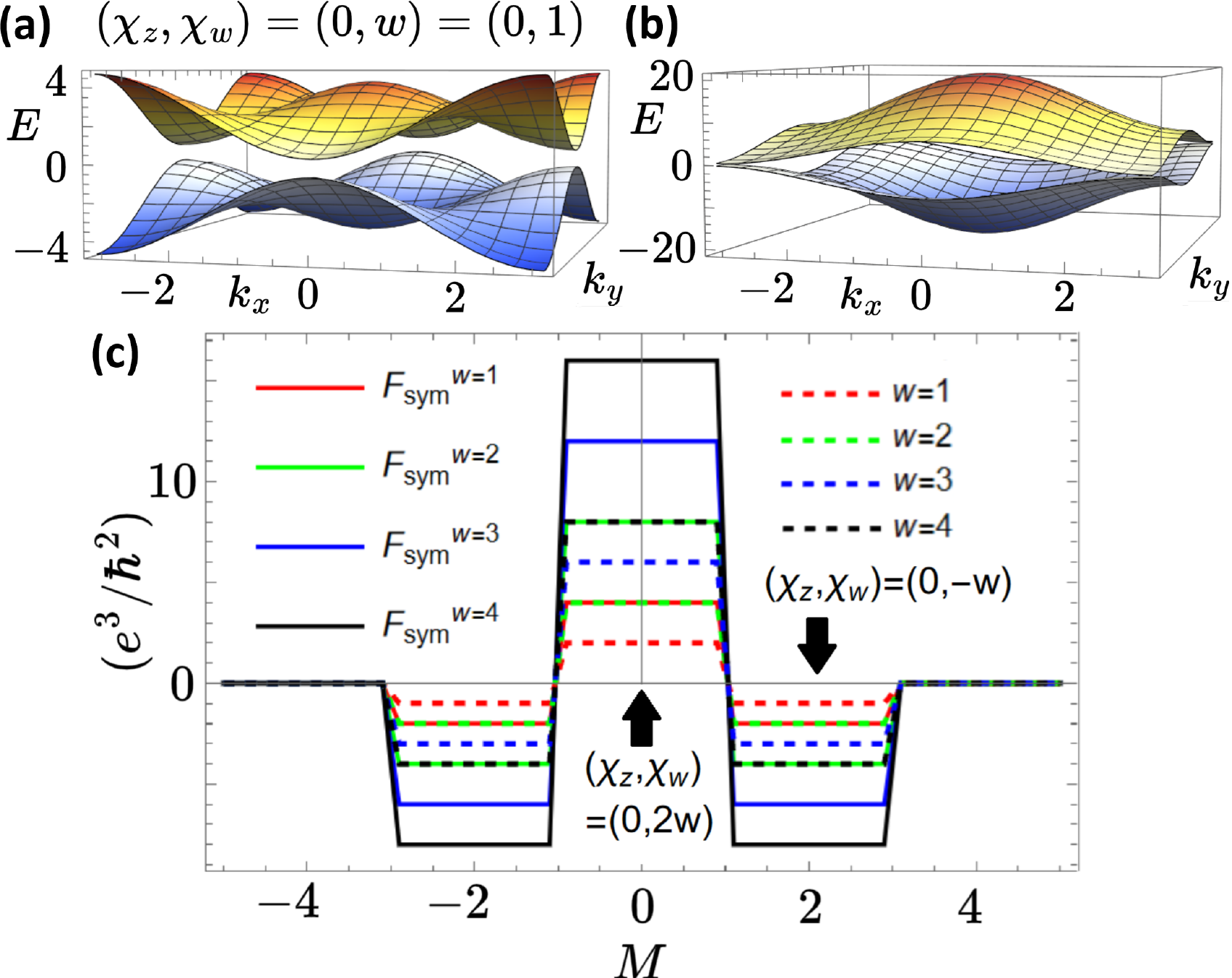}
  \caption{\textbf{(a)} Topological band structure of a real Hopf insulator ($k_z = 0$). \textbf{(b)} Band structure of  RHI model, Eq.~\eqref{eq:Rw_matrix}, for $M = 3/2$, exhibiting  a semimetallic node emerging at the BZ edge $(k_x, k_y, k_z) = (\pi,\pi,0)$, across the topological phase transition. \textbf{(c)} The variation of the quantized shift effect represented by  $F_{\text{sym}}$, indicating TPTs at $M = 3/2$ and $M = 3$, across which both topological bulk invariant (\textit{dashed}) and the associated quantized integrated shift photoconductivities $F_{\text{sym}}$ (\textit{solid}) change sharply. Namely, in the central, low $|M|$ region, $(\chi_z, \chi_w) = (0, 2w)$, and for $3/2 < |M| < 3$, $(\chi_z, \chi_w) = (0, -w)$, with the invariants set by the chosen $w \in \mathbb{Z}$.}
\label{fig:realHopf}
\end{figure}

\section{Torsional representation of symmetrized circular shift photoconductivities}\label{app::C}

As introduced in the main text, the circular shift photoconductivity $\sigma^{cab}_{\text{shift,C}}(\om)$ in terms of the metric connection $C^{mn}_{bca} = (C^{mn}_{bca})^* = \mathfrak{Re}~C^{mn}_{bca}$, under the reality condition (see also Sec.~\ref{app::A}), reads
\beq{eq:c_shift}
   \sigma^{cab}_{\text{shift,C}}(\om) = \frac{e^3}{8\pi^2 \hbar^2} \int \dd^3\kv \sum_{m,n} f_{nm}\mathfrak{Re}~(C^{mn}_{bca}-C^{mn}_{acb}) \delta(\om_{mn}-\om).
\eeq
Now, to derive the torsional representation of circular shift photoconductivities, we start by recognizing that the sum of circular shift photoconductivities can be written as:
\beq{}
\begin{aligned}
     \int \dd \om~ [\sigma^{abc}_{\text{shift,C}}(\om) + \sigma^{bca}_{\text{shift,C}}(\om) + \sigma^{cab}_{\text{shift,C}}(\om)] = \textcolor{white}{ \int \dd \om~ [\sigma^{abc}_{\text{shift,C}}(\om) + \sigma^{bca}_{\text{shift,C}}(\om) + \sigma^{cab}_{\text{shift,C}}(\om)] + 1 - 1 + 1 - 1=}
     \\ \frac{e^3}{8\pi^2 \hbar^2} \int \dd^3\kv \sum_{m,n} f_{nm}\Big( (\mathfrak{Re}~C^{mn}_{cab}-\mathfrak{Re}~C^{mn}_{bac}) + (\mathfrak{Re}~C^{mn}_{abc}-\mathfrak{Re}~C^{mn}_{cba}) + (\mathfrak{Re}~C^{mn}_{bca}-\mathfrak{Re}~C^{mn}_{acb}) \Big) \\= -\frac{e^3}{8\pi^2 \hbar^2} \int \dd^3\kv \sum_{m,n} f_{nm}\Big( (\mathfrak{Re}~C^{mn}_{acb}-\mathfrak{Re}~C^{mn}_{abc}) + (\mathfrak{Re}~C^{mn}_{cba}-\mathfrak{Re}~C^{mn}_{cab}) + (\mathfrak{Re}~C^{mn}_{bac}-\mathfrak{Re}~C^{mn}_{bca}) \Big) \\= -\frac{e^3}{8\pi^2 \hbar^2} \int \dd^3\kv \sum_{m,n} f_{nm} \mathfrak{Re}~(T^{mn}_{acb} + T^{mn}_{cba} + T^{mn}_{bac}),
\end{aligned}
\eeq
where the first equality follows directly from the Eq.~\eqref{eq:c_shift} of the main text on performing the integration over the frequencies, and second equality corresponds identically to reordering the terms. The third equality follows from the definition of the torsion tensor and decomposition of Hermitian connection into real (metric) and imaginary (symplectic) parts. Therefore, the sum of circular shift photoconductivities can be captured purely in terms of the torsion tensor.

\section{Torsion tensor in non-Abelian topological phases satisfying the reality condition}\label{app::D}

First, we rewrite the torsion tensor in terms of the non-Abelian Berry connection, assuming that a symmetry imposing the reality condition is present,
\beq{}
    T^{mn}_{abc} = A^a_{nm} \sum_{m \neq p \neq n} (A^c_{mp} A^b_{pn} - A^b_{mp} A^c_{pn}).
\eeq
Correspondingly, we have:
\beq{}
\begin{split}
    T^{mn}_{abc} + T^{mn}_{bca} + T^{mn}_{cab} =  \sum_{m \neq p \neq n} 
    (A^a_{nm} A^c_{mp} A^b_{pn} -  A^a_{nm} A^b_{mp} A^c_{pn} + A^b_{nm} A^a_{mp} A^c_{pn} +\\ -  A^b_{nm} A^c_{mp} A^a_{pn} + A^c_{nm} A^b_{mp} A^a_{pn} -  A^c_{nm} A^a_{mp} A^b_{pn}),
\end{split}
\eeq
which explicitly reflects the definitional nature of the torsion tensor capturing possible virtual transitions and corresponding matrix elements, as recognized in Ref.~\cite{Ahn_2021rio}, here in terms of the non-Abelian Berry connection elements.

In the subsequent, we evaluate the symmetrized sum of the torsion tensor elements in the four-band and three-band setups of the non-Abelian topological insulators satisfying the reality condition, cf.~Sec.~\ref{app::B}.

\subsection{Torsion in four-band models}

We start with the evaluation of the torsion tensor in the four-band models, relevant for the non-Abelian topologies studied in Refs.~\cite{Lim2023_realhopf, 3plus1}.

\begin{figure*}[t!]
\centering
  \includegraphics[width=\linewidth]{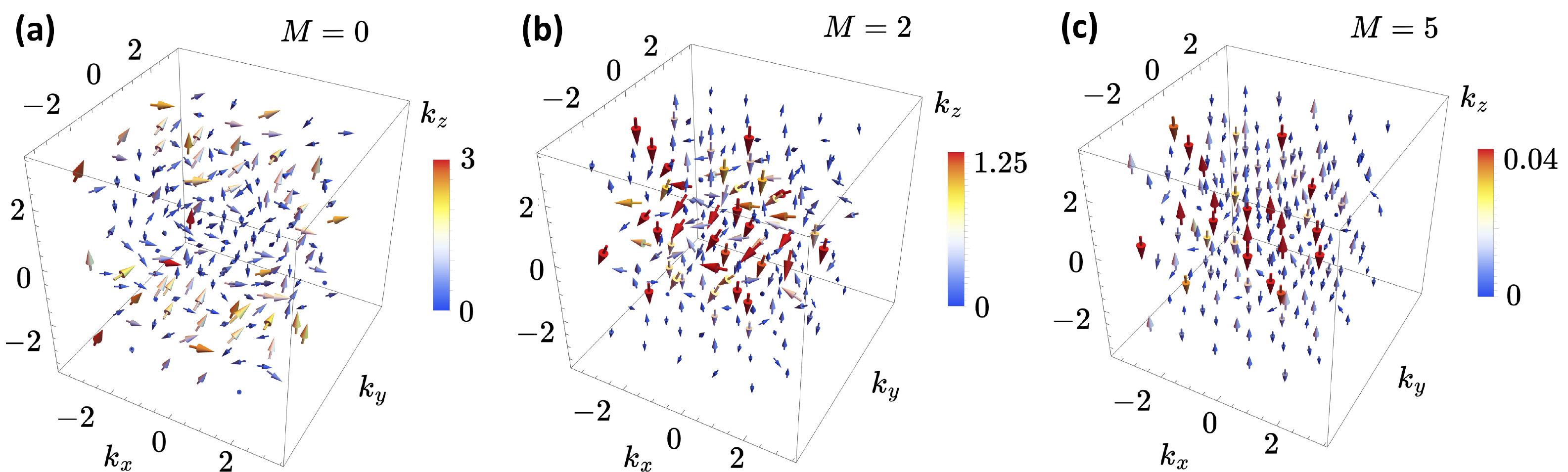}
  \caption{Momentum space fields of the torsion tensor components, represented as vectors ($T_{xyz}, T_{yzx}, T_{zxy}$) over the 3D $\text{BZ} \cong T^3$. We choose the flag phase model studied for the quantized integrated shift effect in the main text, with topological mass $M$ as a free parameter, and $w=1$. \textbf{(a)} Setting $M=0$ yields isoclinic winding numbers $(w_L, w_R) = (-2w, 0) = (-2, 0)$, which is reflected by the high magnitudes of the torsion tensor components, \textbf{(b)} Crossing the TPT at $M=3/2$ reduces the topological indices to $(w_L, w_R) = (w,0)$, and the torsion tensor components individually decrease, yielding a reduced quantized circular shift effect on summation. \textbf{(c)} $M=5$ trivializes the invariants to $(w_L, w_R) = (0,0)$, almost completely removing the torsion tensor components. The leftover values of $T_{xyz}, T_{yzx}, T_{zxy}$ cancel on performing a full summation over the BZ, yielding a vanishing effect quantized by the topological indices $w_L, w_R \in \mathbb{Z}$.}
\label{fig:torsion}
\end{figure*}

Here, we assume that there are 2 occupied bands ($n = 1,2$) and 2 unoccupied bands ($m = 3,4$), obtaining:
\beq{}
\begin{split}
    T^{31}_{abc} + T^{31}_{bca} + T^{31}_{cab} =  
    (A^a_{13} A^c_{32} A^b_{21} -  A^a_{13} A^b_{32} A^c_{21} + A^b_{13} A^a_{32} A^c_{21} -  A^b_{13} A^c_{32} A^a_{21} + A^c_{13} A^b_{32} A^a_{21} -  A^c_{13} A^a_{32} A^b_{21}) +\\+ (A^a_{13} A^c_{34} A^b_{41} -  A^a_{13} A^b_{34} A^c_{41} + A^b_{13} A^a_{34} A^c_{41} -  A^b_{13} A^c_{34} A^a_{41} + A^c_{13} A^b_{34} A^a_{41} -  A^c_{13} A^a_{34} A^b_{41}),
\end{split}
\eeq
\beq{}
\begin{split}
    T^{41}_{abc} + T^{41}_{bca} + T^{41}_{cab} =  
    (A^a_{14} A^c_{42} A^b_{21} -  A^a_{14} A^b_{42} A^c_{21} + A^b_{14} A^a_{42} A^c_{21} -  A^b_{14} A^c_{42} A^a_{21} + A^c_{14} A^b_{42} A^a_{21} -  A^c_{14} A^a_{42} A^b_{21}) +\\+ (A^a_{14} A^c_{43} A^b_{31} -  A^a_{14} A^b_{43} A^c_{31} + A^b_{14} A^a_{43} A^c_{31} -  A^b_{14} A^c_{43} A^a_{31} + A^c_{14} A^b_{43} A^a_{31} -  A^c_{14} A^a_{43} A^b_{31}),
\end{split}
\eeq
\beq{}
\begin{split}
    T^{32}_{abc} + T^{32}_{bca} + T^{32}_{cab} =  
    (A^a_{23} A^c_{31} A^b_{12} -  A^a_{23} A^b_{31} A^c_{12} + A^b_{23} A^a_{31} A^c_{12} -  A^b_{23} A^c_{31} A^a_{12} + A^c_{23} A^b_{31} A^a_{12} -  A^c_{23} A^a_{31} A^b_{12}) +\\+ (A^a_{23} A^c_{34} A^b_{42} -  A^a_{23} A^b_{34} A^c_{42} + A^b_{23} A^a_{34} A^c_{42} -  A^b_{23} A^c_{34} A^a_{42} + A^c_{23} A^b_{34} A^a_{42} -  A^c_{23} A^a_{34} A^b_{42}),
\end{split}
\eeq
\beq{}
\begin{split}
    T^{42}_{abc} + T^{42}_{bca} + T^{42}_{cab} =  
    (A^a_{24} A^c_{41} A^b_{12} -  A^a_{24} A^b_{41} A^c_{12} + A^b_{24} A^a_{41} A^c_{12} -  A^b_{24} A^c_{41} A^a_{12} + A^c_{24} A^b_{41} A^a_{12} -  A^c_{24} A^a_{41} A^b_{12}) +\\+ (A^a_{24} A^c_{34} A^b_{32} -  A^a_{24} A^b_{43} A^c_{32} + A^b_{24} A^a_{43} A^c_{32} -  A^b_{24} A^c_{43} A^a_{32} + A^c_{24} A^b_{43} A^a_{32} -  A^c_{24} A^a_{43} A^b_{32}).
\end{split}
\eeq
On summing all the terms and simplifying using $A^a_{ij}=-A^a_{ji}$ consistently with the reality condition (see Sec.~\ref{app::B}):
\beq{}
\begin{split}
    \sum_{m,n} (T^{mn}_{abc} + T^{mn}_{bca} + T^{mn}_{cab}) = -2A^a_{12}(A^b_{13}A^c_{32}-A^c_{13}A^b_{32}+A^b_{14}A^c_{42}-A^c_{14}A^b_{42}) +\\ -2A^b_{12}(A^c_{13}A^a_{32}-A^a_{13}A^c_{32}+A^c_{14}A^a_{42}-A^a_{14}A^c_{42})
    -2A^c_{12}(A^a_{13}A^b_{32}-A^b_{13}A^a_{32}+A^a_{14}A^b_{42}-A^b_{14}A^a_{42}) +\\
    -2A^a_{34}(A^b_{31}A^c_{14}-A^c_{31}A^b_{14}+A^b_{32}A^c_{24}-A^c_{32}A^b_{24}) +\\ -2A^b_{34}(A^c_{31}A^a_{14}-A^a_{31}A^c_{14}+A^c_{32}A^a_{24}-A^a_{32}A^c_{24})
    -2A^c_{34}(A^a_{31}A^b_{14}-A^b_{31}A^a_{14}+A^a_{32}A^b_{24}-A^b_{32}A^a_{24}).
\end{split}
\eeq
In terms of the two-band Euler connection: $\text{Eu}^{ab}_{nm} = \bra{\partial_a u_n}\ket{\partial_b u_m} - \bra{\partial_b u_n}\ket{\partial_a u_m} = \sum_{n \neq p \neq m} A^a_{np} A^b_{pm} - A^b_{np} A^a_{pm}$, where the second equality follows from resolution of identity $1 = \sum_p \ketbra{u_p}{u_p}$, and the reality condition ($A^a_{ii}=-A^a_{ii}=0$). Hence:
\beq{}
\begin{split}
    \sum_{m,n} (T^{mn}_{abc} + T^{mn}_{bca} + T^{mn}_{cab}) = -2A^a_{12} \text{Eu}^{bc}_{12} - 2A^b_{12} \text{Eu}^{ca}_{12} -2A^c_{12} \text{Eu}^{ab}_{12} - 2A^a_{34} \text{Eu}^{bc}_{34} - 2A^b_{34} \text{Eu}^{ca}_{34} -2A^c_{34} \text{Eu}^{ab}_{34}.
\end{split}
\eeq
On identifying $\textbf{a}^v = \vec{A}_{21} = -\vec{A}_{12}$, $\textbf{a}^c = \vec{A}_{43} = -\vec{A}_{34}$, and finally  $(\text{\textbf{Eu}}^{v/c})_a = \varepsilon_{abc} \text{Eu}^{bc}_{1/3~ 2/4}$, we retrieve:
\beq{}
    \sum_{m,n} (T^{mn}_{abc} + T^{mn}_{bca} + T^{mn}_{cab}) = -2 (\vec{a}^v \cdot \text{\vec{Eu}}^v + \vec{a}^c \cdot \text{\vec{Eu}}^c ).
\eeq
In terms of the language of differential forms:
\beq{}
    \sum_{m,n} (T^{mn}_{abc} + T^{mn}_{bca} + T^{mn}_{cab})~ dk_x \wedge dk_y \wedge dk_z = -2 (a^v \wedge \text{Eu}^v + a^c \wedge \text{Eu}^c), 
\eeq
which is the 3-form integrated over the $\text{BZ} \cong T^3$ to find the sum of the winding, or real Hopf, invariants in the four-band case. Here, the Euler connections (1-forms) and curvatures (2-forms) are defined in terms of the pairs of conduction and valence bands ${a^{v/c} = \bra{u_{2/4}}\dd\ket{u_{1/3}}}$, ${\text{Eu}^{v/c} = \dd a^{v/c}}$. To demonstrate the nontriviality of the 3-form components, we plot the relevant torsion tensor components summed over bands $T_{xyz} \equiv \sum_{m,n} T^{mn}_{xyz}$, $T_{yzx} \equiv \sum_{m,n} T^{mn}_{yzx}$, $T_{zxy} \equiv \sum_{m,n} T^{mn}_{zxy}$ in the form of a vector field $(T_{xyz},T_{yzx},T_{zxy})$ over BZ torus, see Fig.~\ref{fig:torsion}. For the demonstration, we choose the multigap flag phase insulator model from the main text. We notice that significant contributions appear from the edge of BZ. 

The BZ integral of the sum of the torsion tensor components yields quantized numbers reflected by the circular shift photoconductivity, as introduced in the main text. 
It should be stressed again that a four-band bundle, as modelled here in terms of the virtual transitions between four-bands (with amplitudes given by the elements of the non-Abelian Berry connection), can be selectively accessed by tuning the range of frequencies involved in nonlinear optical transitions generating the circular shift current. Moreover, the four-band subspace needs to be isolated from the rest of the valence and conduction bands. Otherwise, the studied homotopy invariants can be trivialized~\cite{bouhon2020geometric}, which emphasizes the multigap nature of these topological invariants. Finally, we notice that if the material is isotropic, then: $\sigma^{xyz}_{\text{shift}}(\om) = \sigma^{yzx}_{\text{shift}}(\om) = \sigma^{zxy}_{\text{shift}}(\om)$. In that case, on combining with the final result of Sec.~\ref{app::C}, the shift photoconductivity integrates to a quantized value for $each$ of the $individual$ directions of light incidence, i.e.~the necessity of combining all symmetrized directions to achieve the quantization condition is removed.

\subsection{Torsion in three-band models}

For completeness, we also introduce the derivation of the symmetrized sums of the torsion tensor elements for 3-band models under the reality condition (Sec.~\ref{app::B}). Now, we assume that there are 2 occupied bands ($n = 1,2$) and 1 unoccupied band ($m = 3$), obtaining:
\beq{}
\begin{split}
    T^{31}_{abc} + T^{31}_{bca} + T^{31}_{cab} =  
    (A^a_{13} A^c_{32} A^b_{21} -  A^a_{13} A^b_{32} A^c_{21} + A^b_{13} A^a_{32} A^c_{21} -  A^b_{13} A^c_{32} A^a_{21} + A^c_{13} A^b_{32} A^a_{21} -  A^c_{13} A^a_{32} A^b_{21}),
\end{split}
\eeq
\beq{}
\begin{split}
    T^{32}_{abc} + T^{32}_{bca} + T^{32}_{cab} =  
    (A^a_{23} A^c_{31} A^b_{12} -  A^a_{23} A^b_{31} A^c_{12} + A^b_{23} A^a_{31} A^c_{12} -  A^b_{23} A^c_{31} A^a_{12} + A^c_{23} A^b_{31} A^a_{12} -  A^c_{23} A^a_{31} A^b_{12}).
\end{split}
\eeq
On summing all the terms and simplifying, using $A^a_{ij}=-A^a_{ji}$:
\beq{}
\begin{split}
    \sum_{m,n} (T^{mn}_{abc} + T^{mn}_{bca} + T^{mn}_{cab}) = -2A^a_{12}(A^b_{13}A^c_{32}-A^c_{13}A^b_{32}) +\\ -2A^b_{12}(A^c_{13}A^a_{32}-A^a_{13}A^c_{32})
    -2A^c_{12}(A^a_{13}A^b_{32}-A^b_{13}A^a_{32}).
\end{split}
\eeq
In terms of the two-band Euler connection: $\text{Eu}^{ab}_{nm} = \bra{\partial_a u_n}\ket{\partial_b u_m} - \bra{\partial_b u_n}\ket{\partial_a u_m} = \sum_{n \neq p \neq m} A^a_{np} A^b_{pm} - A^b_{np} A^a_{pm}$. Hence:
\beq{}
\begin{split}
    \sum_{m,n} (T^{mn}_{abc} + T^{mn}_{bca} + T^{mn}_{cab}) = -2A^a_{12} \text{Eu}^{bc}_{12} - 2A^b_{12} \text{Eu}^{ca}_{12} -2A^c_{12} \text{Eu}^{ab}_{12}.
\end{split}
\eeq
On identifying $\textbf{a}^v = \vec{A}_{21} = -\vec{A}_{12}$, and $(\text{\textbf{Eu}}^v)_a = \varepsilon_{abc} \text{Eu}^{bc}_{12}$, we retrieve:
\beq{}
    \sum_{m,n} (T^{mn}_{abc} + T^{mn}_{bca} + T^{mn}_{cab}) = -2 \vec{a}^v \cdot \text{\vec{Eu}}^v.
\eeq
In terms of differential forms:
\beq{eq:3bandform}
    \sum_{m,n} (T^{mn}_{abc} + T^{mn}_{bca} + T^{mn}_{cab})~ dk_x \wedge dk_y \wedge dk_z = -2 a^v \wedge \text{Eu}^v,
\eeq
which is the 3-form integrated to find the possible quantization in the three-band case. However, contrary to the four-band case, the topology of three-band models in three spatial dimensions has not been studied extensively, as for the time of writing. In particular, in three dimensions, a partitioning of three real bands into 2-band and 1-band subspaces can in principle yield a single $\mathbb{Z}$ invariant, captured by the homotopy $\pi_3[\widetilde{\mathsf{Gr}}_{1,3}(\mathbb{R})] \cong \pi_3[\mathsf{SO}(3)/\mathsf{SO}(2)] \cong \pi_3[S^2] \cong \mathbb{Z}$ associated with a single Hopf index~\cite{Unal_quenched_Euler}. An exhaustive study of such topological phases, supporting the nontrivial real 3-form from Eq.~\eqref{eq:3bandform}, as well as the torsional circular shift photoconductivity response, has also been reported on in Ref.~\cite{jankowski2024nonabelianhopfeulerinsulators}, after the appearance of this work. 

\begin{figure}
\centering
  \includegraphics[width=\columnwidth]{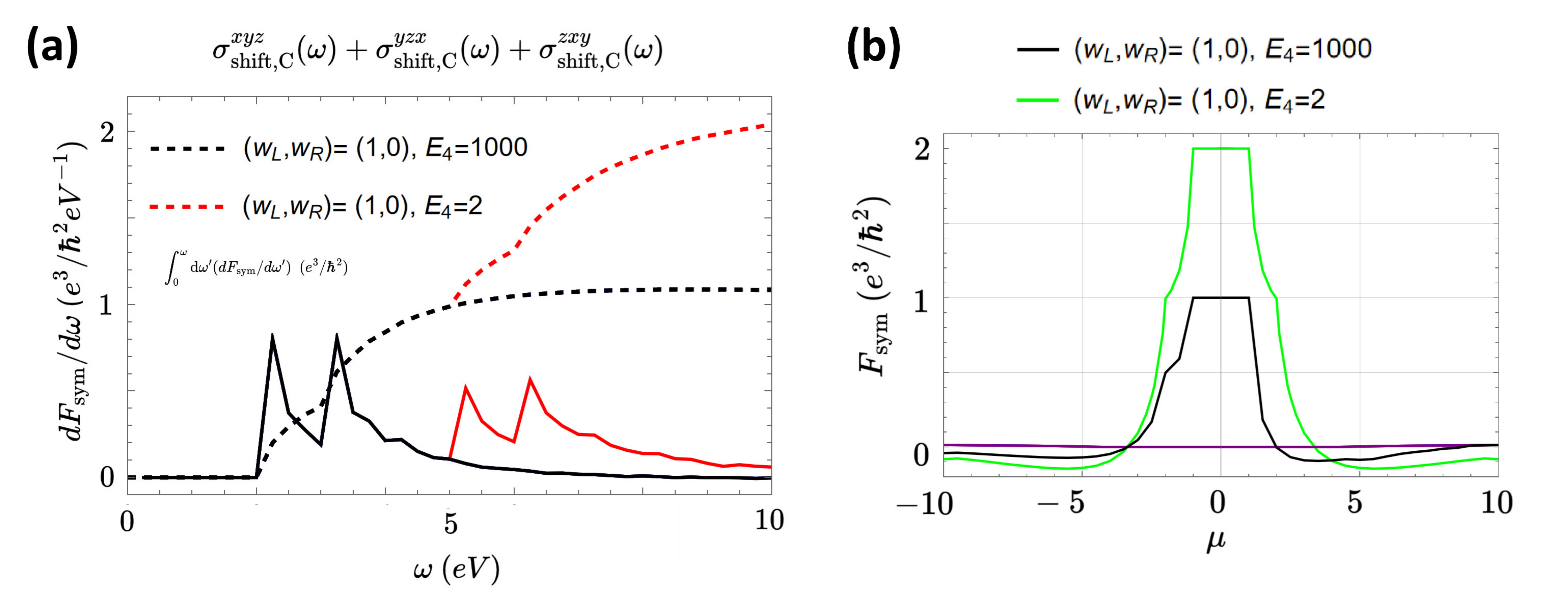}
  \caption{\textbf{(a)} Evolution of the quantized integrated shift response $F_{\text{sym}}$ in the main text models, on sending $E_4 = 2 \rightarrow \infty$. \textbf{(b)}~The change of $F_\text{sym}$ on changing the chemical potential $\mu$, on setting $E_4 = 10^3$ and $\om_{\text{max}} = 200~\text{eV}$ (in the context of the tight-binding models considered in our work,  `eV' is used as an arbitrary unit), which approximates the scenario of sending the uppermost band to infinity in the four-band flag phase models of the main text. We observe that on effectively obtaining a three-band flag phase in that limit, the integrated shift response remains quantized, but halves in magnitude.}
\label{fig:sendInfty}
\end{figure}

Nonetheless, for completeness, we here present how the quantized integrated shift effect evolves on sending one band of the four-band flag phases to infinity, which serves as a proxy for a three-band flag phase. We find that under such scenario, the quantization breaks down into two separate quantized contributions corresponding to the transitions to two individual conduction bands. The evolution of the shift response on sending the energy of one band effectively to infinity is shown in Fig.~\ref{fig:sendInfty}, where the dependence of the integrated response on the chemical potential under such scenario is also included.
\section{Quantized integrated shift responses in real Hopf and Pontryagin-indexed insulators}

Here, we further elaborate on the analogous results for the real Hopf and Pontryagin-indexed insulators, which were mentioned in the main text.

In the case of real Hopf insulators, we utilize the model introduced in Sec.~\ref{app::B} for the calculations of quantized integrated shift response in RHIs, as demonstrated in Fig.~\ref{fig:realHopf}. These results are analogous to the calculations for flag phases that were demonstrated in the main text. Namely, in Fig.~\ref{fig:realHopf}, we present the evolution of quantized shift responses in the RHI models with $(\chi_z, \chi_w) = (0,w)$, generated with a matrix of Bloch eigenvectors,
\beq{eq:Rw_matrix}
\begin{split}
    R_w(\kv) = \sin wk_x~\Gamma_{11} + \ii \sin k_y~\Gamma_{23} - \sin k_z~\Gamma_{31} +\\+ (M - \cos wk_x - \cos k_y - \cos k_z) \Gamma_{03}.
\end{split}
\eeq
Here, $\Gamma_{ij} = \sigma_i \otimes \sigma_j$ are $4 \cross 4$ Kronecker products of identity and Pauli matrices $(\sigma_0, \sigma_1, \sigma_2, \sigma_3) \equiv (\mathbf{1}_2, \sigma_x, \sigma_y, \sigma_z)$. Additionally, here, $M$ is the topological mass, changes of which can cause the phase transitions trivializing the RHI.

Next, we elaborate in the single-index quantization by the $Q$ invariant of the Pontryagin-indexed insulators. Here, to access all transitions necessary for the quantization of the response, one out of three occupied bands below the gap between the one-band and three-band subspaces needs to become depopulated upon doping. Correspondingly, as mentioned in the main text, such non-Abelian phases enjoy a simple quantization by $Q \in \mathbb{Z}$, and in that case the quantized shift response, can be analogously realized on photoexcitations to two unoccupied bands, although upon further doping and band engineering.


Namely, the Pontryagin-indexed insulators can be doped, such that only two bands are occupied, as the Fermi level is set at the band crossings in dispersive limit, which can be further gapped by an infinitesimal perturbation, as these nodes are not protected by the invariant $Q$. Gapping out the nodes obtains a phase classified by another flag manifold $\mathsf{Fl}_{1,1,2}$~\cite{jankowski2024nonabelianhopfeulerinsulators}, with homotopy $\pi_3[\mathsf{Fl}_{1,1,2}] \cong \mathbb{Z}^2$, which supports two indices $w_-, w_+ \in \mathbb{Z}$, such that $Q = w_- + w_+$. This effectively results in a similar quantization condition $F_{\text{sym}} = \frac{2e^3}{\hbar}(w_- + w_+)  = \frac{2e^3}{\hbar} Q$, intermediate between the fully-gapped flag phases, and the real Hopf insulators with both two-band subspaces hosting band degeneracies.

For the step-by-step derivation of the quantized shift responses, as captured by the real Chern-Simons forms and the torsion tensors, we refer to Secs.~\ref{app::C},~\ref{app::D}.

\section{Quantization of the shift effect in the presence of additional bands}

In this section, we elaborate on the effects of additional bands on the presence of the quantization of the integrated shift effect. As mentioned in the main text, we find that the quantization of $F_\text{sym}$ is robust, as long as the four-band subspace does not hybridize with the additional bands to trivialize the topological invariant, and the appropriate frequency cutoff $\om_{\text{max}}$ is provided. Here, necessarily, the gaps between the additional bands and the topological four-band subspace are preserved over the BZ.

\begin{figure}
\centering
  \includegraphics[width=\columnwidth]{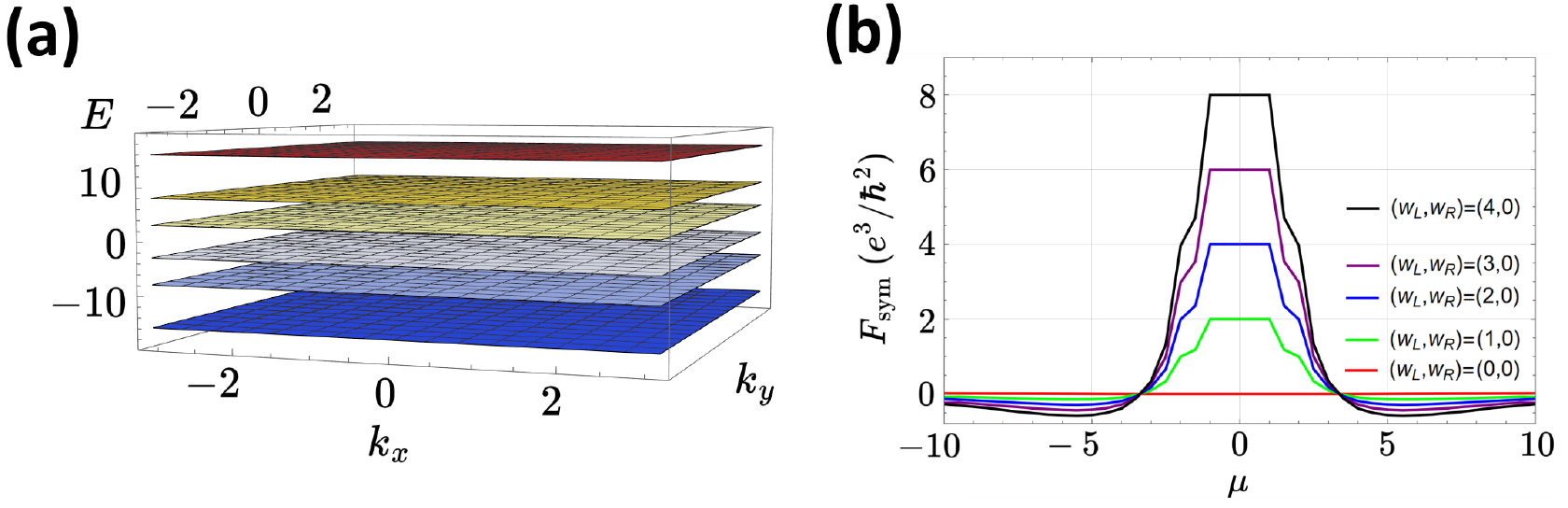}
  \caption{ \textbf{(a)} A band structure of flag phase realizing a four-band multigap topology, with two additional topologically-trivial bands at $E_0 = -15$ and $E_5 = 15$. \textbf{(b)}~The change of $F_\text{sym}$ on moving the chemical potential $\mu$ in the presence of additional valence and conduction bands. The chemical potential modifies the occupation of six bands. We observe that the topological quantization persists despite the occupation of a trivial band, as long as the topological invariant of the four-band subspace is preserved and nontrivial, while the additional bands contribute no torsion.}
\label{fig:addBands}
\end{figure}

{\it Adding trivial bands:} First, we consider adding trivial bands to the system, which is evidently important to many real systems, i.e. physical materials. Let us first assume adding a single band with energy above and a single band with energy below to the topological four-band subspace. Correspondingly, these bands carry non-vanishing transition dipole moments (being the non-Abelian connection~\cite{Ahn_2021rio}), which are constrained by the non-hybridization condition, see Fig.~\ref{fig:addBands}. Such a scenario can be modelled with a Hamiltonian,
\beq{}
    H(\textbf{k}) = V_{\text{add}}(\textbf{k})~\text{diag} [E_0, E_1, E_2, E_3, E_4, E_5]~V_{\text{add}}(\textbf{k})^{\text{T}},
\eeq
with $E_0 < E_1 < E_2 < E_3 < E_4 < E_5$, where $V_{\text{add}}$ involves the original four topological bands, as e.g. in the model realizations of the main text, and additonal bands $\ket{u_{5,\kv}}, \ket{u_{0,\kv}}$. In Fig.~\ref{fig:addBands}, we for example consider

\beq{}
\ket{u_{5,\kv}} = \Big( \cos (k_x + k_y + k_z), 0, 0, 0, 0, -\sin (k_x + k_y + k_z) \Big)^\text{T},
\eeq
\beq{}
\ket{u_{0,\kv}} = \Big( \sin (k_x + k_y + k_z), 0, 0, 0, 0, \cos (k_x + k_y + k_z) \Big)^\text{T}.
\eeq

If the additional bands do not hybridize with the topological bands, i.e. the coefficients of orbitals constituting the topological bands vanish in their case, naturally preserving the multigap invariants, we observe that $F_{\text{sym}}$ remains quantized, and the integrated shift response demonstrates a similar dependence on the chemical potential $\mu$ (on~setting $\om_{\text{max}} = \infty$), as the four-band Hamiltonian with no additional bands, see Fig.~\ref{fig:addBands}. In particular, the fact that there are no corrections to $F_{\text{sym}}$ due to the additional two bands can be understood as following from the fact that two bands that are coupled to each other, but not hybridized with any other bands, cannot carry any torsion~\cite{AhnPRX}. Hence, here, the torsion tensor elements of the additional bands, e.g. $T^{50}_{xyz}$, vanish identically, and no additional contributions to $F_{\text{sym}}$ emerge. 

We now consider adding more than two bands satisfying the imposed $\mathcal{PT}$ symmetry. Correspondingly, we address different scenarios on increasing the frequency from $\om_{\text{max}}$, which originally targeted the four-band subspace only. On increasing the $\omega_{\text{max}}$ to entail additional \textit{full} bands, we note that even if the additional bands host nontrivial torsion tensor components locally in $\textbf{k}$-space, the integrated shift effect $F_{\text{sym}}$ will remain similarly quantized when $\omega_{\text{max}}$ allows for excitations from the additional bands over the entire BZ. In that case, the 
{\it integral over the entire} BZ
of the additional torsion tensor contributions needs to vanish,
\beq{}
    \int_\text{BZ} \dd^3\kv \sum_{n,m \neq 1,2,3,4} (T^{mn}_{xyz} + T^{mn}_{yzx} + T^{mn}_{zxy}) = 0,
\eeq
if the added bands are trivial, i.e. they host no additional multigap topological invariants. The reason that this condition holds is that the torsion tensor integral analogously corresponds to the real Chern-Simons 3-forms hosted by the additional bands. An integral of any Chern-Simons form yields only (topologically) quantized values. Therefore, an additional nontrivial value acquired on the integration over the entire BZ would mean that the bands hosted some nontrivial topology yielding a nontrivial quantized (winding) number due to a topological Chern-Simons form, contrary to the assumption that the additional bands are trivial. We stress that such scenario {\it is analogous to quantized circular dichroism in two-dimensions}, with the Berry curvature 2-forms in place of Chern-Simons 3-forms. Namely, on increasing the frequency cutoff, any additional trivial bands involved in the optical transitions are expected to \text{not} contribute to the dichroism due to their trivial Chern number, unlike the topological bands, which in terms of the optical transition dipole matrix elements (with which the Berry curvature form can be rewritten) yield nontrivial contributions to the summed dichroic response~\cite{doi:10.1126/sciadv.1701207, Asteria_2019}.

As a limiting case, we recognize that on increasing the cutoff frequency $\om_{\text{max}}$, such that range of the photon frequencies up to $\om_{\text{max}}$ is able to excite to (or from) the additional bands \textit{partially}, i.e. only a three-dimensional patch $\mathcal{D} \in \text{BZ}$ is on resonance and becomes excited, that generically yields,

\beq{}
    \int_\mathcal{\mathcal{D}} \dd^3\kv \sum_{n,m \neq 1,2,3,4} (T^{mn}_{xyz} + T^{mn}_{yzx} + T^{mn}_{zxy}) \neq 0.
\eeq

In that particular case, the quantization condition could naturally break down due to the non-vanishing (partial) torsional contributions of additional bands, under the increased modified cutoff $\om_{\text{max}}$. However, it should be stressed that such scenario, that loses the quantization, can be naturally omitted, if the range of frequencies up to $\om_{\text{max}}$ was \textit{reduced} to the original range admitting the photoexcitations only in the low-energy topological four-band subspace, \textit{or} was increased such that the whole bands can be excited, which is more natural, e.g. in cold-atom experiments~\cite{doi:10.1126/sciadv.1701207, Asteria_2019}. We moreover stress that under the assumption of \textit{partial} excitations between the bands, the well-known quantized circular dichroisms also break down. That is, if one probes only a part of a disperse Chern band with photoexcitations, there is no definite quantization, as similarly the optical weights do not add up to a quantized value.

{\it Adding topological bands:} The final scenario that can emerge is for $\om_{\text{max}}$ accessing a higher gap, such that $F_{\text{sym}}$ remains quantized, but changes its value due to the addition of further \textit{topological} bands (and according transitions) with non-vanishing torsion that is beyond the contribution of the initial four-band subspace which enjoys a quantization by the multigap invariants (e.g. $w_L, w_R$). From the above formulae, it is clear that under the non-hybridization condition, this will need to involve the addition of more than two bands, as otherwise the contributions will identically vanish. Such a scenario can be achieved by embedding the four-band system characterized by invariants ($w_L, w_R$) in the principal gap, i.e. the middle gap, of another flag phase Hamiltonian with positive and negative band energies higher in magnitude, and invariants ($w'_L, w'_R$). Under such circumstances, the additional bands modify the quantized sum as:
\beq{}
    F_{\text{sym}} = \frac{2e^3}{\hbar^2} (w_L + w_R) \rightarrow \frac{2e^3}{\hbar^2} (w_L + w_R + w'_L + w'_R),
\eeq
as soon as $\om_\text{max}$ becomes sufficient to set all the transitions involving bands with the invariants $(w'_L, w'_R)$ on resonance.

Finally, and most importantly, we notice that the described scenarios (of removing and recovering the quantization on setting the increasing $\om_\text{max}$ in the higher bands and gaps) could alternate up to the point when $\om_{\text{max}}$ crosses the ionization threshold. Reaching that threshold with $\om_{\text{max}}$ corresponds to involving an infinite number of higher bands admitting photoexcitations over the full BZ. Consistently with the previous arguments (about Chern-Simons 3-forms), this topologically requires the quantized ultimate value of $F_{\text{sym}}$, while the continuum of free particle states above the ionization threshold contributes no further torsion. Hence, the final integrated response in the limit $\om_{\text{max}} \rightarrow \infty$ (as mentioned in the main text) is topologically quantized, directly paralleling the quantized magnetic circular dichroism that can be measured on sending an upper limit $\om_{\text{max}} \rightarrow \infty$~\cite{doi:10.1126/sciadv.1701207}. In that analogy, the proposed quantized integrated shift effect should be measurable from second-order nonlinear dc responses, on similarly setting an infinite cutoff ($\om_{\text{max}} \rightarrow \infty$), as in the spirit of the quantized circular dichroism measurements~\cite{doi:10.1126/sciadv.1701207, Asteria_2019}, which, on the contrary, involve a response at linear order.


{\it As concluding remarks:} We stress that while the contributions of \textit{any} additional bands can be naturally eliminated, if their corresponding photoexcitations require $\om > \om_{\text{max}}$, this means that the no-hybridization condition is sufficient on its own, and can be in principle realized in real electronic materials under an interplay of orbital symmetries and selection rules that need to be obeyed. As long as the selection rules in the system dictate the transition dipole matrix elements $r^i_{mn}$ between the topological bands (which could be constituted by, e.g. only $d$-orbitals) and the other bands to vanish $(r^i_{mn} = 0)$, no additional torsion can be contributed to the spectral shift integral $F_{\text{sym}}$ by the other non-topological bands \textit{as long as $\om_\text{max}$ spans an energy window exciting full bands}. This condition on $\om_\text{max}$ manifestly preserves the quantization of $F_{\text{sym}}$, despite the presence of additional bands, and their transitions being possibly on resonance. Finally, in the limit $\om_\text{max} \rightarrow \infty$, the quantization under the no-hybridization condition is also retrieved, as it corresponds to exciting to an infinite number of $full$ bands, i.e. all $\textit{k}$-points in the BZ are targeted. As mentioned, this scenario parallels with the \textit{infinite} frequency cutoff case of the quantized circular dichroism in the two-dimensional Chern insulators~\cite{doi:10.1126/sciadv.1701207, Asteria_2019}.

%
\clearpage

\section*{Supplementary References}
\bibliographystyle{apsrev4-1}
\bibliography{supp_references.bib}